\documentclass[amsmath,amssymb,aps,twocolumn,longbibliography,notitlepage]{revtex4-1}
\usepackage{graphicx}
\usepackage{verbatim}
\usepackage{subfigure}
\usepackage{mathrsfs}
\usepackage[letterpaper=true,colorlinks=false]{hyperref}
\usepackage{color}

\newcommand{\beq}{\begin{equation}}
\newcommand{\eeq}{\end{equation}}

\definecolor{correction}{RGB}{255,0,0}

\begin{document}

\title{BaH molecular spectroscopy with relevance to laser cooling}
\author{M. G. Tarallo, G. Z. Iwata, T. Zelevinsky}
\affiliation{Department of Physics, Columbia University, 538 West 120th Street, New York, NY 10027-5255, USA}
\date{\today}

\begin{abstract}
We describe a simple experimental apparatus for laser ablation of barium monohydride (BaH) molecules and the study of their rovibrational spectra that are relevant to direct laser cooling.  We present a detailed analysis of the properties of ablation plumes that can improve the understanding of surface ablation and deposition technologies.  A range of absorption spectroscopy and collisional thermalization regimes has been studied. We directly measured the Franck-Condon factor of the $\mathrm{B}^2\Sigma^+(v'=0)\leftarrow\mathrm{X}^2\Sigma^+(v''=1)$ transition. Prospects for production of a high luminosity cryogenic BaH beam are outlined.  This molecule is a promising candidate for laser cooling and ultracold fragmentation, both of which are precursors to novel experiments in many-body physics and precision measurement.
\end{abstract}

\maketitle

\section{Introduction}
\label{sec:Intro}

The study of the spectra of diatomic molecules has a longstanding importance in many research areas from physical chemistry to astrophysics and nuclear physics \cite{Herzberg}. Identifying and characterizing the level structure of these molecules with high precision is crucial for expanding the scope of applications for these physically rich systems.  Many diatomic species of interest, because of their high reactivity, must be produced from gaseous or solid precursors, or probed via astronomical observations.  Among these, the light diatomic hydrides, and in particular alkaline-earth-metal hydrides, are of great importance for the understanding of chemical bonding because of their amenability to theoretical treatment.  These molecules are also ideal candidates for direct laser cooling \cite{DiRosa} toward production of polar molecules near absolute zero \cite{Carr}, as well as for ultracold fragmentation for improved precision measurements with hydrogen isotopes, due to the favorably large Ba$/$H mass ratio \cite{LanePRA15_HFromBaH}.

\begin{figure}
	\centering
	\includegraphics[width=0.48\textwidth]{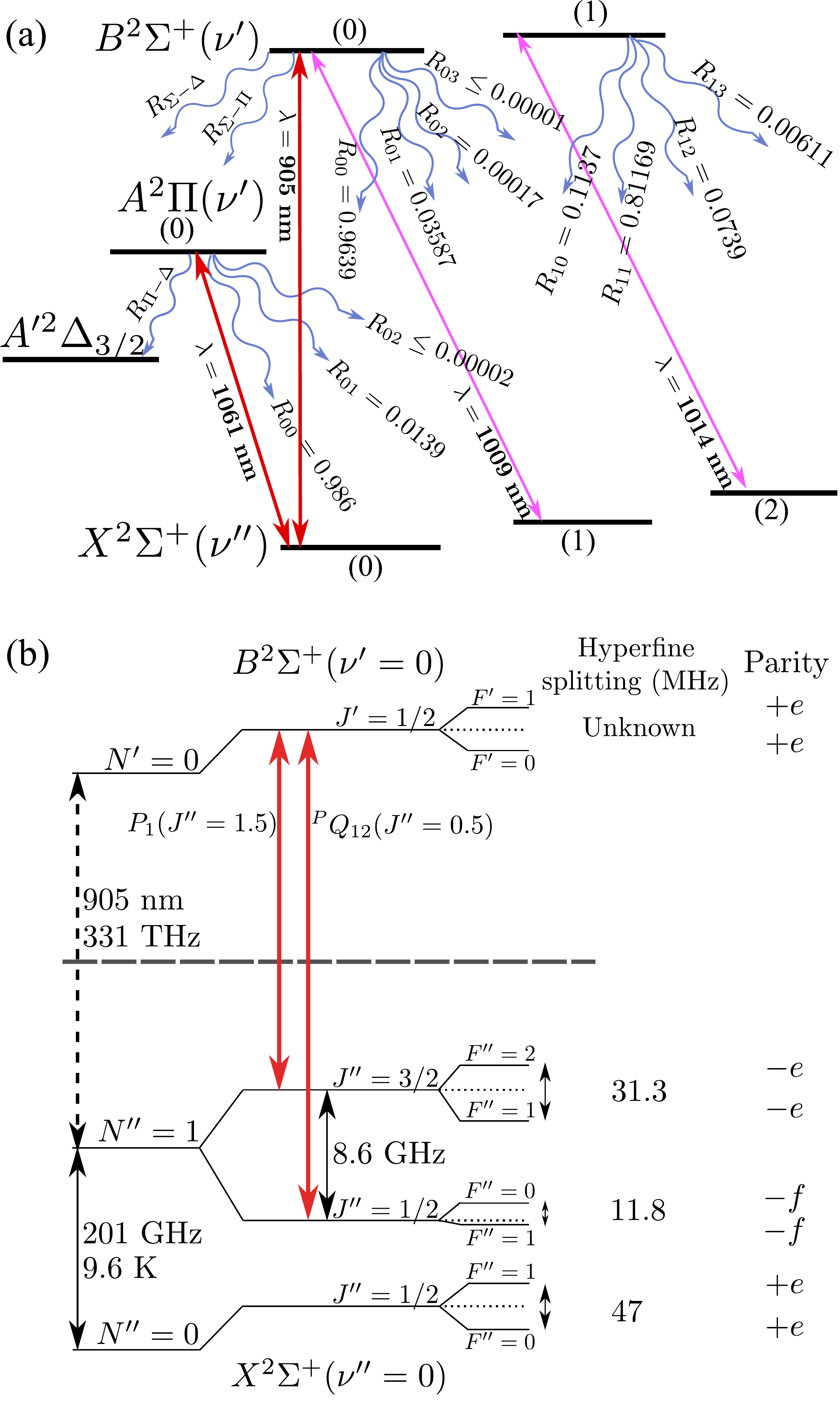}
	\caption{(a) Electronic and vibrational level structure for two possible laser cooling transitions in BaH.  Straight lines represent laser driven transitions, while wavy lines represent radiative decay.  $R_{v''v'}$ are the relative transition moment magnitudes.  (b) Rotational and spin structure of the $\mathrm{B}^2\Sigma^+(v'=0)-\mathrm{X}^2\Sigma^+(v''=0)$ manifold that is relevant to laser cooling.}
	\label{fig:lasercooling}
\end{figure}
In this work we study the production and the spectral properties of barium monohydride (BaH) that are relevant to laser cooling.
Figure \ref{fig:lasercooling} shows the electronic, rovibrational, and spin structure important for molecular laser cooling.
Since 1932, studies have been performed on the BaH electronic ground state, $\mathrm{X}^2\Sigma^+$, and seven excited electronic states \cite{fredwatson,watson,kopp,Appel85,barrow,Bernath}.  The two lowest excited electronic states, $\mathrm{A}^2\Pi$ and $\mathrm{B}^2\Sigma^+$, lie in the near infrared spectral region, and both can be accessed by commercially available diode lasers.  Furthermore, they both exhibit highly diagonal decay to the ground state \cite{rama,DiRosa}, which is crucial for laser cooling.  In fact, both the $\mathrm{A}^2\Pi$ and $\mathrm{B}^2\Sigma^+$ states may enjoy complementary laser cooling and trapping advantages:  the former is predicted to have more favorable decay branching ratios \cite{LanePRA15_HFromBaH}, while the latter has a large magnetic moment that should ensure much stronger magneto-optical trapping forces \cite{tarbMOT}.  Laser cooling has been attempted with $\mathrm{A}^2\Pi$ molecular states \cite{shuman,hummon,zhelya}, but not with $\mathrm{B}^2\Sigma^+$.  To this end, we focus here on the $\mathrm{B}^2\Sigma^+$ state of BaH from the point of view of laser cooling.
In previous studies, BaH was mainly produced in hydrogen-filled furnaces \cite{Appel85}.  Here we present a simple approach to precise spectroscopy of BaH molecules, as well as many other diatomic species, that is directly applicable to buffer-gas cooled molecular beams \cite{DeMilleBarryPCCP11_CryogenicMolecularBeams,DoyleHutzlerCR12_BufferGasBeams} that are used for loading molecular magneto-optical traps (MOTs) \cite{Barry,hummon}.  The BaH gas is obtained from a solid BaH$_2$ precursor by pulsed laser ablation in a small vacuum chamber constructed from standard components.

The paper is organized as follows. In Sec. \ref{sec:setup} we describe the experimental setup used to produce and probe BaH dimers.  In Sec. \ref{sec:theo} we model the time- and frequency-resolved laser absorption of the ablation plume with and without a buffer gas.  Next, the experimental results are shown in Sec. \ref{sec:results}, where we also characterize the molecular yield and thermal properties as functions of the ablation laser fluence.  Finally, in Sec. \ref{sec:outlook} we discuss the considerations and implications for laser cooling BaH molecules.

\section{Laser ablation and spectroscopy setup}
\label{sec:setup}

\begin{figure}
\centering
\includegraphics[width=.48\textwidth]{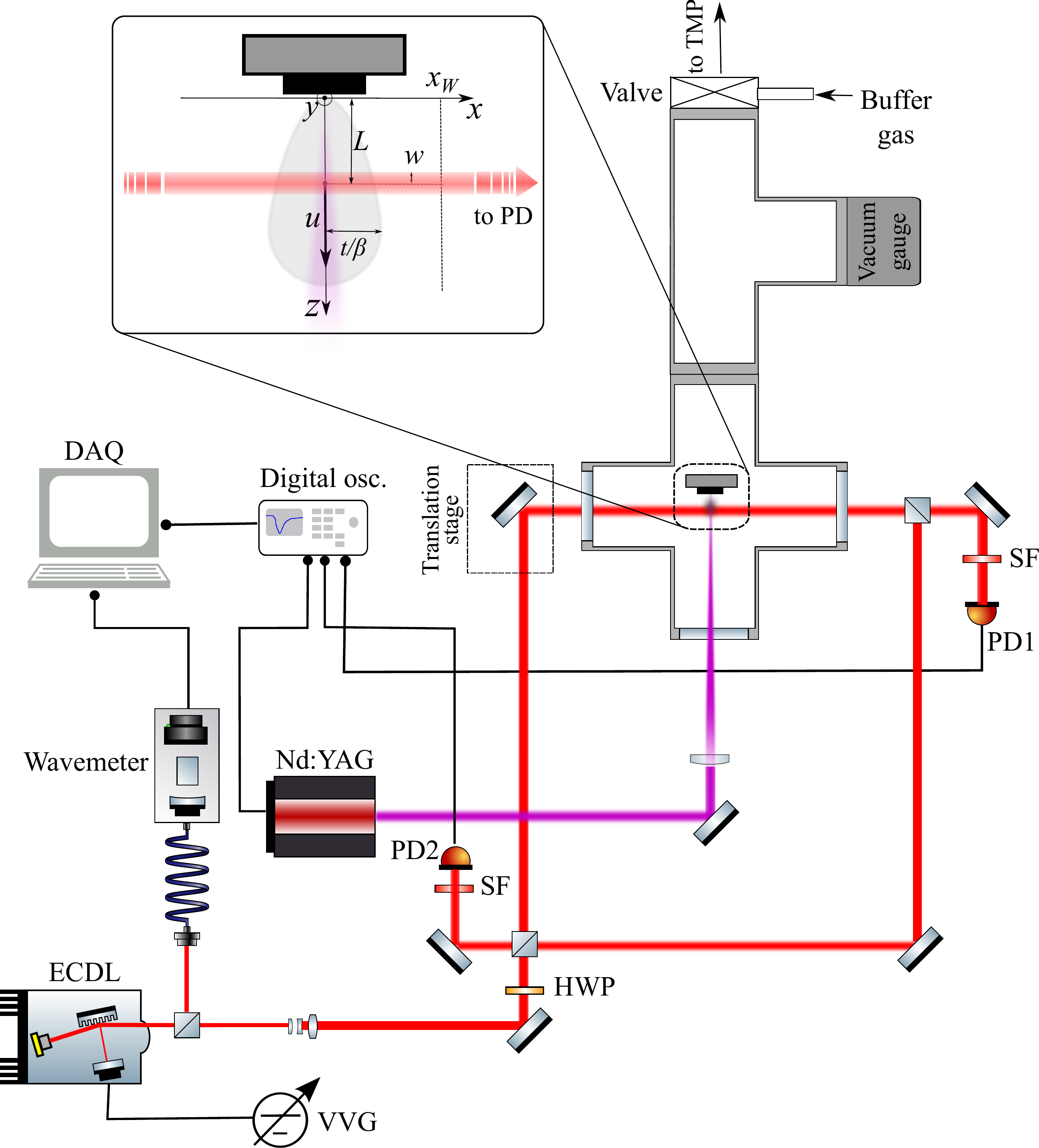}
\caption{Schematic drawing of the ablation test chamber and spectroscopy setup. HWP: half-wave plate; PD: photodetector; SF: spectral filter; TMP: turbo-molecular pump; VVG: variable voltage generator.  The inset illustrates the experimental geometry in the proximity to the target and the resulting ablation plume with a forward speed $u$.  The probe beam is nearly normal to the plume.  $x_W$:  position of the chamber window.}
\label{fig:setup}
\end{figure}
The experimental apparatus is sketched in Fig.~\ref{fig:setup}. The ablation chamber consists of a standard four-way cross with three optical windows and a port leading to the vacuum pump.  The focused Nd:YAG ablation laser (BigSky Ultra) light at 1064 nm with 50 mJ$/$pulse enters through the central window and hits the target glued to the holder at the center of the cross, and the resulting ablation plume is probed along the perpendicular axis.  The energy, or fluence, of the ablation light impinging on the target can be varied by adjusting the focusing lens position.  Probe light is produced by an extended-cavity diode laser (ECDL) in a Littman-Metcalf configuration, tunable between 904 and 914 nm.  We perform precise laser absorption spectroscopy (LAS) on the ablated plume of molecules by monitoring the probe laser extinction level after ablation.  Beam splitters allow for counterpropagating two-beam LAS for precise measurements of the Doppler shift.

The probe light is detected by a fast photodiode (NewFocus 1801) with a narrow-band (10 nm) interference filter, which prevents ablation fluorescence from saturating the detector.  Absorption signals are then recorded by a digital oscilloscope with a response time of 5 ns, which is triggered by the ablation laser Q-switch clock signal.  The recorded time-of-flight (TOF) absorption profile allows us to measure the time evolution of the molecule density across the probe beam in a chosen rotational state.  Precise probe wavelength control is achieved with a 0.01 pm resolution HighFinesse wavemeter.  The setup allows for a controlled variation of many experimental parameters including the ablation beam focus, ablation angle, probe beam size, and distance between the probe beam and the target.

The molecular targets are commercially obtained solid pieces of BaH$_2$ with 99\% purity that we choose for their flat faces and ease of gluing to the holder.  We have also performed ablation spectroscopy on atomic Yb and Sr for comparison and calibration purposes.  Finally, buffer gas can be introduced into the chamber through an inlet just before the vacuum pump valve, and the background and buffer gas pressure are monitored by a full range vacuum gauge (Pirani / cold cathode).

\section{Modeling absorption spectroscopy with ablated molecules}
\label{sec:theo}

Resonant LAS is a straightforward method to characterize laser ablation \cite{harnafi,cheung,geohegan,bushaw}.  Starting from a basic model to describe the molecular ablation plume, in this section we discuss the time- and frequency-resolved LAS signals to characterize the physical properties of the molecular yield and study its internal energy spectra.

\subsection{Time-resolved resonant absorption by the ablation plume}
\label{sec:PlumeAbsorption}

\begin{figure}
	\centering
	\includegraphics[width=0.48\textwidth]{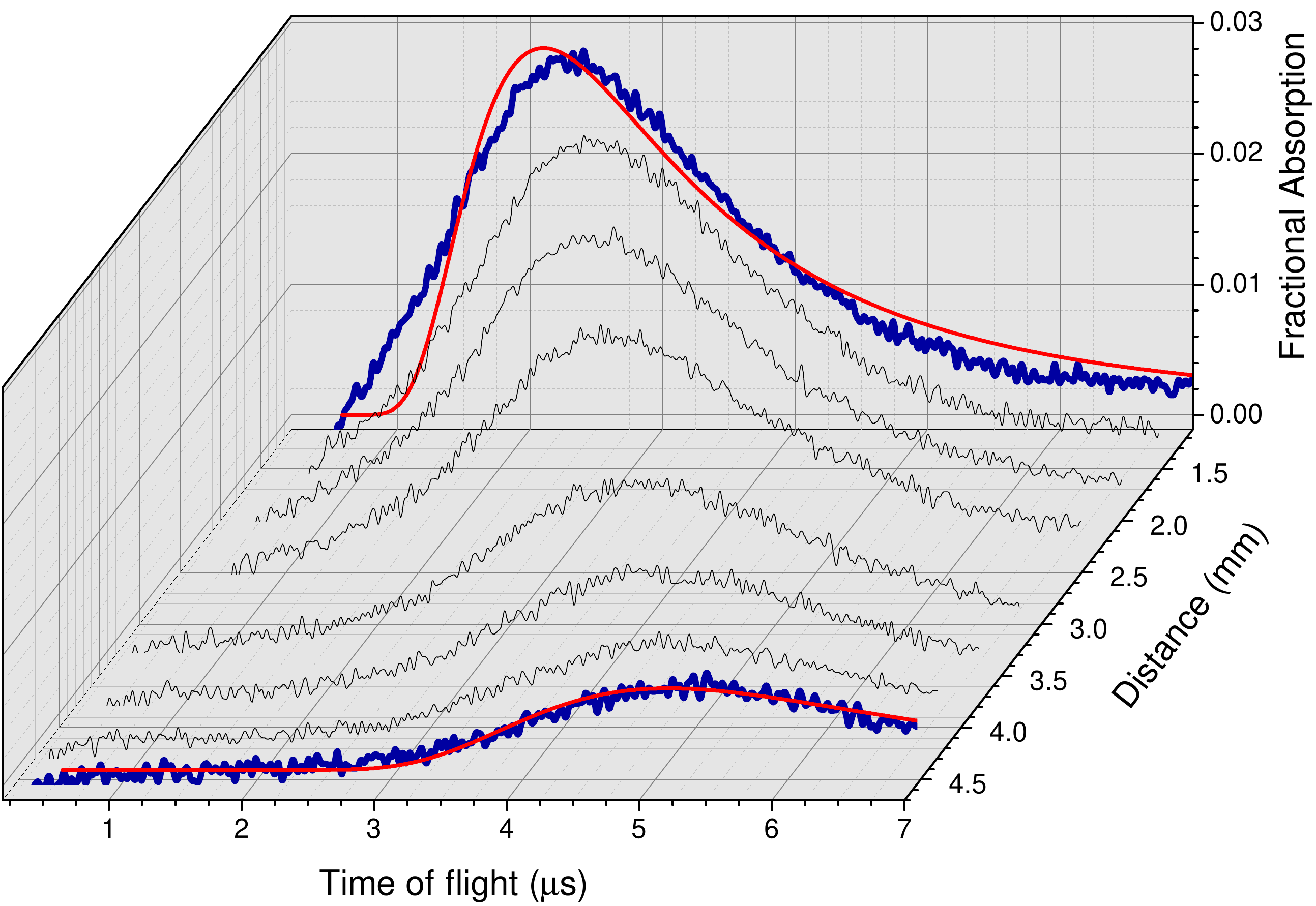}
	\caption{TOF absorption signals of BaH ablation plume recorded at different distances $L$ of the probe laser from the target surface.}
	\label{fig:fig2}
\end{figure}
An example of time-resolved absorption due to the transit of an ablation plume through the probe beam is shown in Fig. \ref{fig:fig2}.  We clearly observe a dependence of the absorption peak shape and position on the probe distance $L$.  The peak shift yields the center-of-mass molecular speed with respect to the line of sight.  By modeling the observed signal we reconstruct the fundamental kinetic quantities that characterize the molecular ablation process:  temperature, forward velocity, and the initial density.

The processes leading to pulsed-laser generated gas clouds have been extensively studied and modeled for thin film deposition \cite{PLD,Belouet}.  We begin with a widely accepted hydrodynamical model \cite{Kools}  which consists of an initial elliptical gas cloud generated by desorption from a surface plasma that rapidly ($t_{\mathrm{eq}}\lesssim100$ ns) equilibrates to a full Maxwell-Boltzmann distribution.  In the coordinate system defined in Fig. \ref{fig:setup}, this is described by
\begin{eqnarray}\label{eq:SMB}
p(\vec{v},T,u)= \frac{\beta^3}{\sqrt{\pi^3}}\exp \left[-\beta^2(v_x^2+v_y^2+(v_z-u)^2)\right]
\label{eq:MBBeamDistrib}
\end{eqnarray}
which is moving with a stream velocity $\vec{u}$ normal to the target surface. In Eq. (\ref{eq:MBBeamDistrib}), $\beta=\sqrt{M/(2k_BT)}\equiv v_{\mathrm{th}}^{-1}$ is the inverse of the most probable thermal velocity, $M$ is the molecular mass, and $T$ is the temperature after the equilibration time $t_{\mathrm{eq}}$.  The effective temperature of an ablated diatomic molecule is $T\sim0.8T_s$ \cite{Kelly}, where $T_s$ is the ablation surface temperature.

The collisionless free expansion of an atomic or molecular gas can be described by solving the Boltzmann equation for the phase-space distribution $f(\vec{x},\vec{v},t)=f_0(\vec{x}-\vec{v}t,\vec{v})$, where $f_0$ is the phase-space density at time $t=0$.  For a rapidly equilibrating ejection process, $f_0(\vec{x}-\vec{v}t,\vec{v})=p(\vec{v},T,u)n_0(\vec{x}-\vec{v}t)$, where $n_0(\vec{x})$ is the initial spatial distribution. Because of the Gaussian shape of the pulsed (ablation) laser beam profile, the initial density distribution is a two-dimensional Gaussian function with radius $w_{\mathrm{pl}}$.  We find the density function for a freely expanding gas governed by a shifted Maxwell-Boltzmann distribution (\ref{eq:SMB}) via $n(\vec{x},t)=\int d\vec{v}p(\vec{v},T,u)n_0(\vec{x}-\vec{v}t)$.
The finite ablation area introduces a time scale $t_{\mathrm{eq}}=w_{\mathrm{pl}}\beta\approx200$ ns for the expanding gas to fully equilibrate (assuming $T=2000$ K, $w_{\mathrm{pl}}=100$ $\mu$m).  For $t\gg t_{\mathrm{eq}}$, the density function $n(\vec{x},t)\propto t^{-3} \exp{\left[-\beta^2(x^2+y^2+(z-ut)^2)/t^2)\right]}$ exhibits a $t^{-3}$ asymptotic behavior as predicted in Ref. \cite{Kools}.

From the time dependence of the plume density, we can calculate the probe absorption. Since the probe laser with a Gaussian width $w$ and intensity $I$ crosses the ablation plume perpendicularly, the absorption signal is a TOF density measurement integrated over the $x$ transverse direction. The Beer-Lambert law, $-dI/I\approx n(\vec{x},t)\sigma(\omega)dx$, where $\sigma(\omega)$ is the frequency-dependent light-matter interaction cross section, describes the absorption in the unsaturated regime.
To calculate the photodetector signal given by the change in probe laser power $P_0$, we integrate over $y$ and $z$, which results in the time-resolved fractional absorption
\begin{eqnarray}
\frac{\Delta P(t)}{P_0}&\equiv&\mathcal{A}(t)
=1 \\
\nonumber &-&\frac{\sigma(\omega)}{P_0}\int_{-\infty}^{\infty}\int_{-\infty}^{\infty}\int_0^{\infty}n(\vec{x},t)I(y,z)dx\,dy\,dz,
\label{eq:mathADef}
\end{eqnarray}
obtaining
\begin{eqnarray}
\mathcal{A}(t)
\approx\frac{N_0\sigma(\omega)\beta^2}{\pi t^2}e^{-\frac{(L-t u)^2\beta^2}{t^2}},
\label{eq:mathA}
\end{eqnarray}
where we have simplified the full expression for $L>w/\sqrt{2}$, $t>\beta w$.  Note that $\mathcal{A}(t\rightarrow\infty)\propto1/t^2$, implying that TOF absorption signals fall off slower than fluorescence signals ($\propto1/t^3$) \cite{geohegan}.
Equation (\ref{eq:mathA}) can be used to fit the TOF profiles as in Fig. \ref{fig:fig2}.
The molecular yield can be estimated from the peak of the fractional absorption,
\beq\label{eq:yield}
N_0\approx\mathcal{A}(t_p) \frac{e^{\frac{(L-t_p u)^2\beta^2}{t_p^2}} \pi  t_p^2}{\beta ^2 \sigma(\omega) },
\eeq
where $t_p$ is the time at which the fractional absorption reaches its peak value. Equivalently, the yield can be also expressed in terms of the initial density by considering that $N_0\approx\pi n_0 v_{\mathrm{th}} w_{\mathrm{pl}}^2 \delta t_p/2$, where the emission time $\delta t_p$ is assumed to be on the order of the laser pulse width \cite{harnafi}.


\subsection{Frequency dependent absorption and saturation effects}
\label{sec:sat}

The fractional absorption at $t_p$ is proportional to the cross section $\sigma(\omega)$. In addition to the natural linewidth, the frequency dependence of $\sigma(\omega)$ must include the Doppler shift of the probe laser frequency with respect to the moving particles. Integration of Eq. (\ref{eq:mathADef})  over the velocity distribution (\ref{eq:SMB}) yields a fractional absorption spectrum \cite{georginov}
\beq\label{eq:freqspec}
\mathcal{A}(\omega)\propto\sigma(\omega)\equiv\sigma_{lu}^0\frac{c\beta}{\omega_0\sqrt{\pi }} e^{\displaystyle{{-\frac{\beta ^2 [c (\omega -\omega_0)+u\, \omega_0 \sin\theta]^2}{\omega_0^2}} }},
\eeq
which is the Gaussian lineshape of the absorption cross section with a full width at half maximum $\Delta\omega = \sqrt{8\ln(2)k_BT\omega_0^2/(Mc^2)}$ and a Doppler shifted center due to the misalignment angle $\theta$ of the line-of-sight relative to the normal.  Here, $\omega_0=2\pi c/\lambda_0$ is the angular frequency of the optical transition resonance, $M$ is the molecular mass, and $c$ is the speed of light.  Because of the $\Delta\omega(T)$ dependence, a measurement of the absorption spectrum provides a direct determination of the transverse translational temperature of the ablation plume. 

We have introduced the integrated cross section
\[
\sigma_{ul}^0 = \frac{g_u}{g_l}\frac{\lambda_0^2}{4}A_{ul},
\]
where $A_{ul}$ is the spontaneous emission rate from the upper to the lower molecular level, and $g_u,\,g_l$ are the respective degeneracies.  Hence a measurement of the absorption cross section at resonance, $\sigma(\omega_0)=\sqrt{8\ln2}\sigma_{ul}^0/\Delta\omega$, provides information on the spontaneous decay process occurring in the electronic transition. This would require an \emph{a priori} knowledge of the ablation molecular yield, which fluctuates from shot to shot. However, the strength of the $\mathrm{B}-\mathrm{X}\;(0,0)$ band can be estimated by measuring the saturation intensity at which non-linear absorption takes place~\cite{Wall}.

The standard saturation intensity is
\beq
I_s =\frac{g_l}{g_u}\,\frac{\pi h c}{\lambda_0^3\,\tau}\,\frac{1}{r(J'',N'')\,R_{00}}\quad,
\eeq
where $\tau$ = 125(2) ns is the lifetime of the $B\,^2\Sigma(v=0)$ state~\cite{berg}, $R_{00}$ is the vibrational branching ratio which is proportional to the Frank-Condon factor $q_{00}$, and $r(J'',N'')$ is the rotational branching ratio for the dipole transition moment of the excited rotational level $|N',J'\rangle$ to a particular rotational level $|N'',J''\rangle$ of the $X$ state.  The rotational branching ratio can be expressed as function of the H\"oln-London factor $\mathscr{S}_{J'N'J''N''}$ as~\cite{Watson20085}
\beq
r(J'',N'') = \frac{\mathscr{S}_{J'N'J''N''}}{(2 J'+1)}.
\eeq

For transient laser absorption by an ablation plume, the reduction of the cross section due to Doppler broadening and the transit time $\tau_t=2w/v_p$ taken by the density peak to pass through the probe laser must be included in the calculation. Saturation effects can be modeled by solving the rate equations for the level populations in the time interval $\Delta t \sim 2\tau_t$ centered on the peak of the temporal absorption signal, in which the total number of particles remains constant~\cite{demtr}.

The rotational state $|\mathrm{B},J'=0.5, N'=0\rangle$ forms a quasi-closed three-level system with the two rotational levels $N''=1$ of the first two vibrational bands in the electronic ground state X, since $R_{00}+R_{01}\approx1$. In this case, the effective saturation intensity is (Fig. \ref{fig:lasercooling})
\beq\label{eq:Isat}
I_{s}^{00} \simeq I_s\frac{1.2\,\Gamma_t\Delta\omega}{\Gamma^2}\frac{\Gamma+\Gamma_t}{[1-r(J'',N'')R_{01}]\Gamma+3\Gamma_t},
\eeq
where $\Gamma_t  =1/\tau_t$ is the transit time broadening and $\Gamma=1/\tau$ is the total radiative relaxation rate.

Another method to measure the excited state branching ratios is to introduce a second laser to probe the $\mathrm{B}­\mathrm{X}\;(v=0,1)$ transition.  Its fractional absorption $\mathcal{A}_{01}$ will depend on the presence of the driving laser on the $\mathrm{B}-\mathrm{X}\;(0,0)$ transition.  For low intensities, we can approximate the fractional absorption from Eq. (\ref{eq:yield}) as $\mathcal{A}_{00}\approx n(t_p)\sigma_{00}(\omega_{00})\Delta x(1­s_0)$.  While stand­alone absorption on the (0,1) band depends on the population of the first excited vibrational state, the differential absorption due to the presence of the $\mathrm{B}-\mathrm{X}(0,0)$ driving laser can be related to all accessible quantities as
\beq\label{eq:Adiff}
\Delta\mathcal{A}_{01} = \mathcal{A}_{00}^0\chi(\Gamma,\Gamma_t,f_1,R_{01})s_0\left(\frac{\lambda_{01}}{\lambda_{00}}\right)^3\frac{R_{01}}{1-R_{01}},
\eeq
where $\mathcal{A}_{00}^0$ is the absorption of the (0,0) laser without the additional probe beam, and
$$
\chi =1+\frac{f_1 [(1-r) \Gamma +5 {\Gamma_t}]-5 {\Gamma_t}}{(1-{f_1}) [(1-{r} {R_{01}}) \Gamma +3 {\Gamma_t}]}
$$
is a function of the initial population fraction in the first excited vibrational level, $f_1$, which takes on positive values between 0.4 and 1.

Equations (\ref{eq:Isat},\ref{eq:Adiff}) provide two independent experimental methods to estimate the vibrational branching ratio $R_{01}=q_{01}(\omega_{01}/\omega_{00})^3$, and thus the vibronic Franck-Condon factor $q_{01}$, which is a prerequisite for direct laser cooling.

\section{Results}
\label{sec:results}

\subsection{Thermal properties and molecular yield of the ablation plume}
\label{sec:ablation}

\begin{figure}
	\centering
	\subfigure{
	\includegraphics[width=0.48\textwidth]{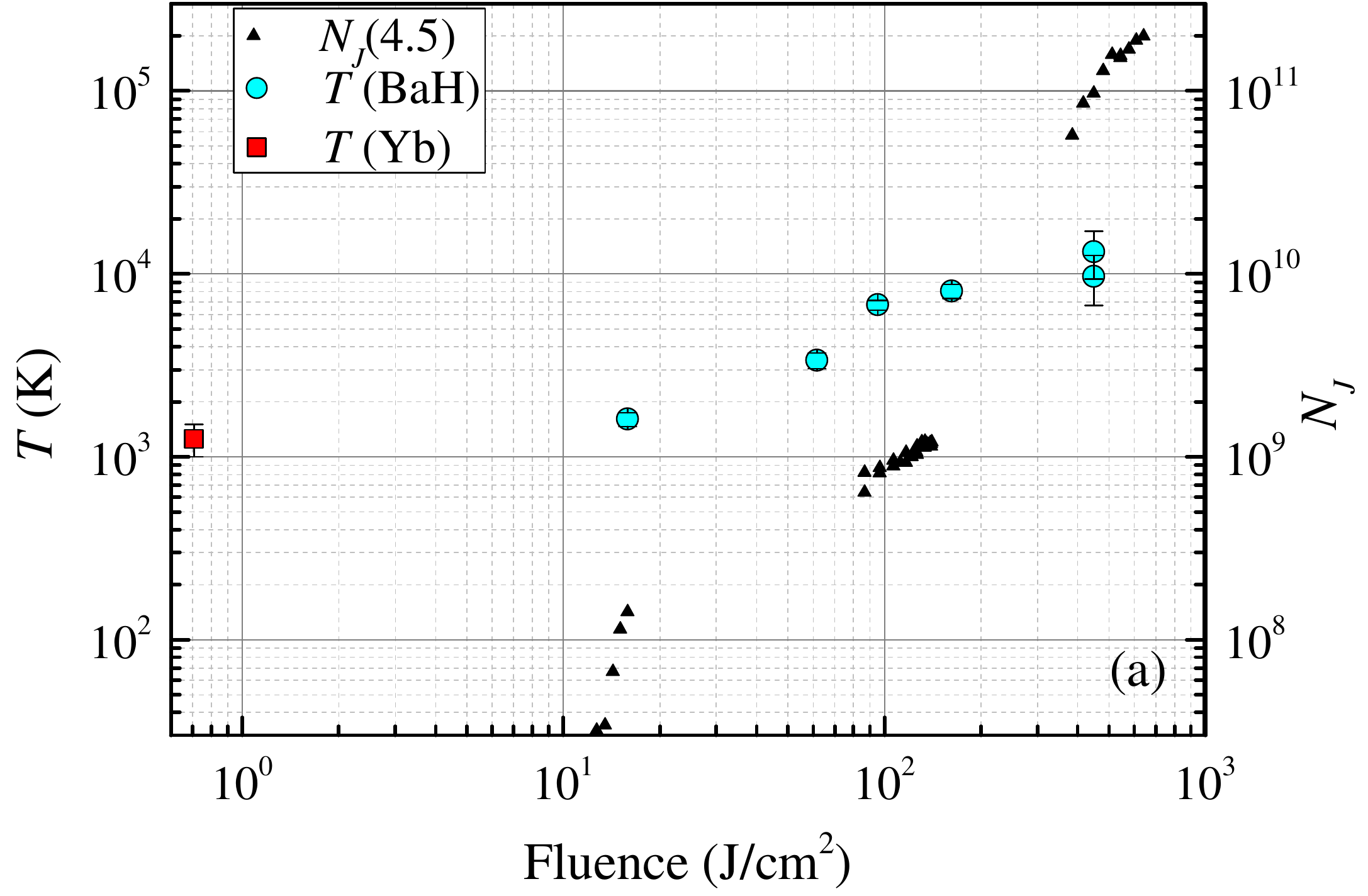}}
	~~~
	\subfigure{
	\includegraphics[width=0.4\textwidth]{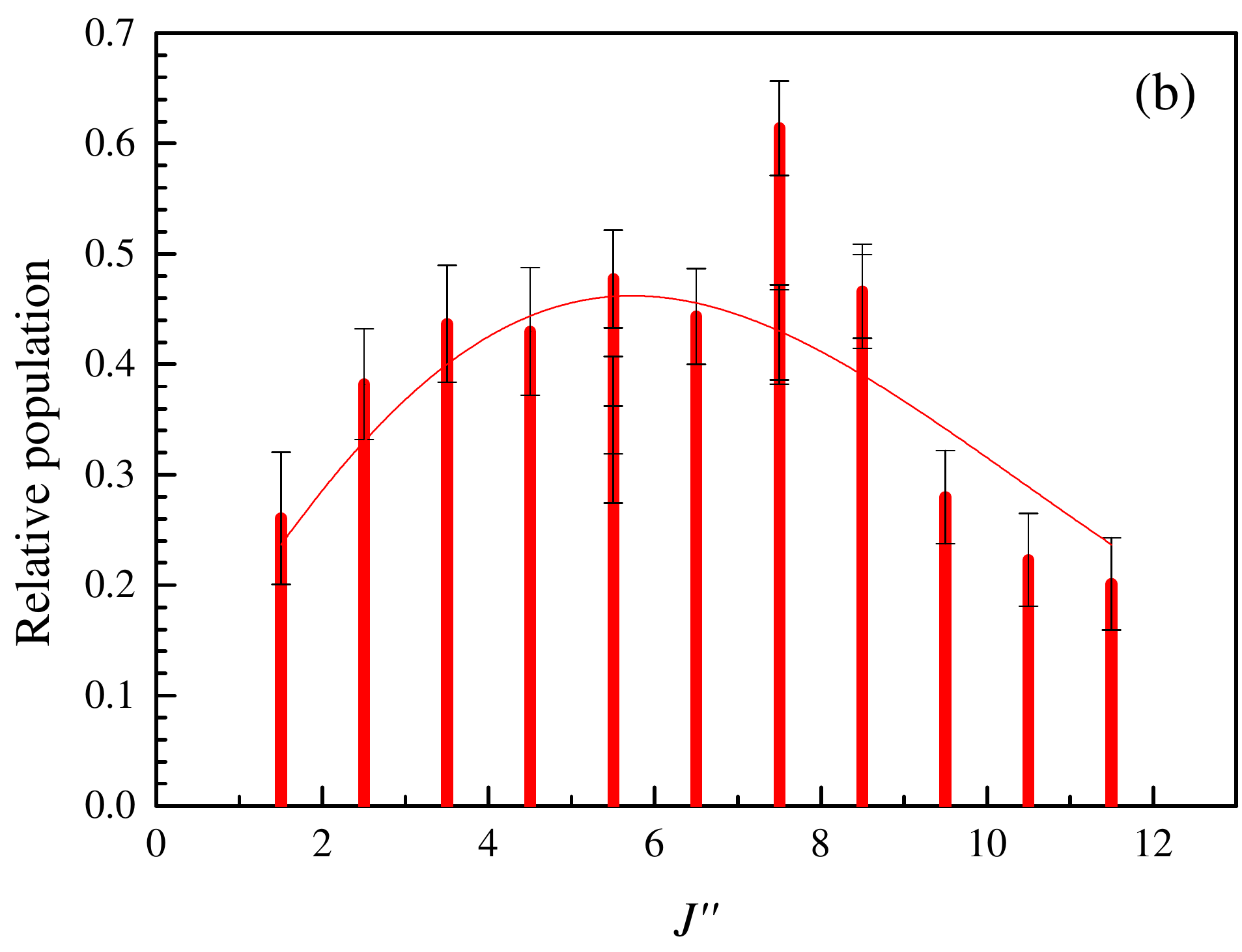}}	
	\caption{Experimental characterization of the molecular temperature via frequency-resolved LAS. (a) Translational temperature and molecular yield as functions of the ablation laser fluence.  (b) Relative populations of the $P_2$ rotational states.}
	\label{fig:yield}
\end{figure}
Measurements of the BaH molecular yield and translational temperature are shown in Fig.~\ref{fig:yield}(a) as a function of the 1064 nm pulsed ablation laser fluence.  The probe laser was tuned to the $P_2(4.5)$ transition (see Sec.~\ref{sec:LinNon}).  We clearly observe the ablation threshold around 10 J$/$cm$^2$ at which the $|N=5,J=4.5\rangle$ state population is $\sim10^7$.  The ablation threshold depends on the physical properties of the target surface such as hardness, roughness, and reflectivity.  Both the hardness and the density of our BaH$_2$ chunks are low ($\lesssim3$ g/cm$^3$) compared, for instance, to metallic Yb or hot pressed targets.  Hence we expect a large potential for improvement by switching to hard targets of BaH$_2$ (4 g/cm$^3$ \cite{AmericanElementsPrivate}) as done in Ref.~\cite{Lu} for CaH molecules.  We also observed a dependence of the molecular yield both on the pulsed laser repetition rate $f_r$ and on the number of shots on the target.  For a laser fluence of 16 J/cm$^2$, the initial yield is nearly constant from $f_r$ = 1 Hz to 15 Hz, and then diminishes, while the shot-to-shot degradation rate depends linearly on $f_r$, with a characteristic decay constant $\tau$($f_r=2$ Hz) $\approx2000$ shots.  Surprisingly, at $f_r=15$ Hz, after $\sim100$ laser pulses and the associated decay of the yield, the signal increased to roughly twice its starting strength.  We interpret this as the constant heating of the target due to a weak thermal link to its environment.  At $f_r=20$ Hz, we also observed a two-peak absorption profile due to thermal desorption of the hot surface after the first ablation event.

The translational temperatures of the ablated molecules versus fluence were recorded by measuring the Doppler widths of absorption profiles.
The minimum temperature of $\sim1800$ K was measured at 16 J/cm$^2$, corresponding to a molecular yield of $1.4\times10^8$ and a forward velocity $u\approx910$ m/s.  For higher fluence we observed a linear increase of the translational temperature.  The extrapolated temperature at the ablation threshold is $\sim1500$ K.  This estimate is in fair agreement with previous experiments on BaH gases created in furnaces with hydrogen atmosphere, which indicated that BaH$_2$ has a decomposition point near 1300 K \cite{Magg}.

The distribution of the total population among the rotational ground states $|N'',J''\rangle$ is determined from the relative strengths of observed absorption lines.  These strengths are related to the Maxwell-Boltzmann distribution via the rotational temperature $T_r$~\cite{Herzberg}.  A measurement of the rotational temperature using the $P_2$ branch is shown in Fig.~\ref{fig:yield}(b).
The best fit yields $T_r\approx400$ K, one order of magnitude lower than the translational temperature.  Discrepancies between translational and rotational temperatures in ablation plumes were have been reported~\cite{rottemp}, and confirm the non-thermal origin of ablation.  By using this rotational temperature, we can estimate the total molecular yield as
\[
N_{\mathrm{tot}}\approx N_J\frac{(2S+1)k_BT_r}{(2J+1)\,hcB_X}\exp\left[\frac{hcF(N)}{k_bT_r}\right] \sim 3\times10^{9},
\]
where the effective rotational spacing $F(N)=B_XN^2+(B_X-\gamma_X/2)N-\gamma_X/2$ takes into account the spin splitting of the transition branch.

The total molecular yield we obtained is similar to that of other alkaline-earth hydride beam experiments \cite{Campbell,DoyleWeinsteinNature98_mKCaH}.  For a single-stage cryogenic buffer gas cell the maximum experimental extraction efficiency in the hydrodynamic regime is $\sim50\%$ \cite{DoyleHutzlerCR12_BufferGasBeams}, thus $N_{\mathrm{tot}}\sim1.5\times10^9$ molecules/pulse.

\subsection{Diffusion and cooling with room temperature buffer gas}
\label{sec:buffer}

We have studied the possibility to reduce Doppler broadening of the molecular spectra, and at the same time conducted an initial investigation of the BaH collision cross sections with noble gases, by filling the ablation chamber with buffer gas at room temperature ($T_{\mathrm{bg}}=297$ K).  The buffer gas dissipates both the translational and rotational energies.  Translational cooling relies on elastic collisions, but inelastic scattering may also occur for rotational thermalization.  This can explain the small difference between rotational and translational temperatures measured in cryogenic buffer-gas molecular beams~\cite{Hutzler}.

\subsubsection{Translational cooling}
\label{sec:TranslationalCooling}

The dynamics of an ablation plume produced inside a background gas has been studied in a variety of systems. It can be divided into two phases:  an initial ballistic expansion of the plume away from the target, followed by diffusion which depends inversely on the buffer gas pressure \cite{skoff}.  During both phases, thermalization of the plume with the buffer gas occurs \cite{Campbell}.  The translational thermalization with an inert buffer gas has been modeled by assuming elastic collisions between two hard spheres of  mass $m$ (buffer gas) and $M$ (ablated species).  In a hard-sphere model, from energy and momentum conservation the average temperature change per collision event is $\Delta T =-\kappa (T-T_{\mathrm{bg}})$, where $\kappa\equiv(M+m)^2/(Mm)$.  This implies the ablation plume temperature evolution \cite{deCarv}
\beq\label{eq:buffcooling}
\frac{dT}{dt}=-R(T-T_{\mathrm{bg}})/\kappa,
\eeq
where $R=n_{\mathrm{bg}}\sigma_{\mathrm{BaH},\mathrm{bg}}\bar{v}$ is the elastic collision rate determined by the buffer gas density $n_{\mathrm{bg}}=p/(k_BT_{\mathrm{bg}})$ with $p$ being the gas pressure, the thermally averaged collision cross section $\sigma_{\mathrm{BaH},\mathrm{bg}}$, and the average relative speed
$$
\bar{v} = \sqrt{\frac{8 k_B}{\pi}\left(\frac{T}{M}+\frac{T_{\mathrm{bg}}}{m}\right)}\sim \sqrt{\frac{8 k_B T}{\pi \mu}}.
$$

\begin{figure}
	\centering
	\includegraphics[width=0.48\textwidth]{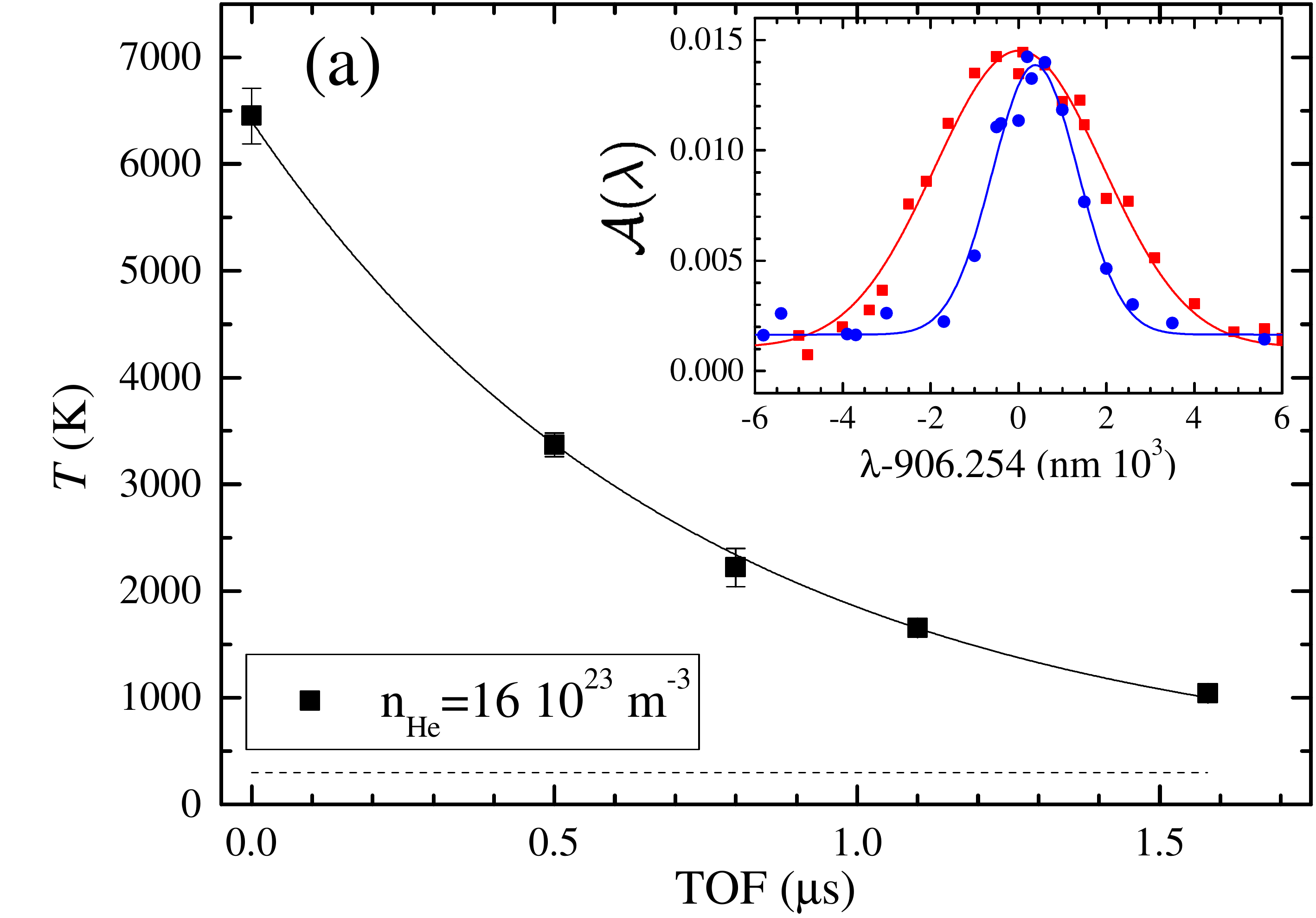}
	\includegraphics[width=0.48\textwidth]{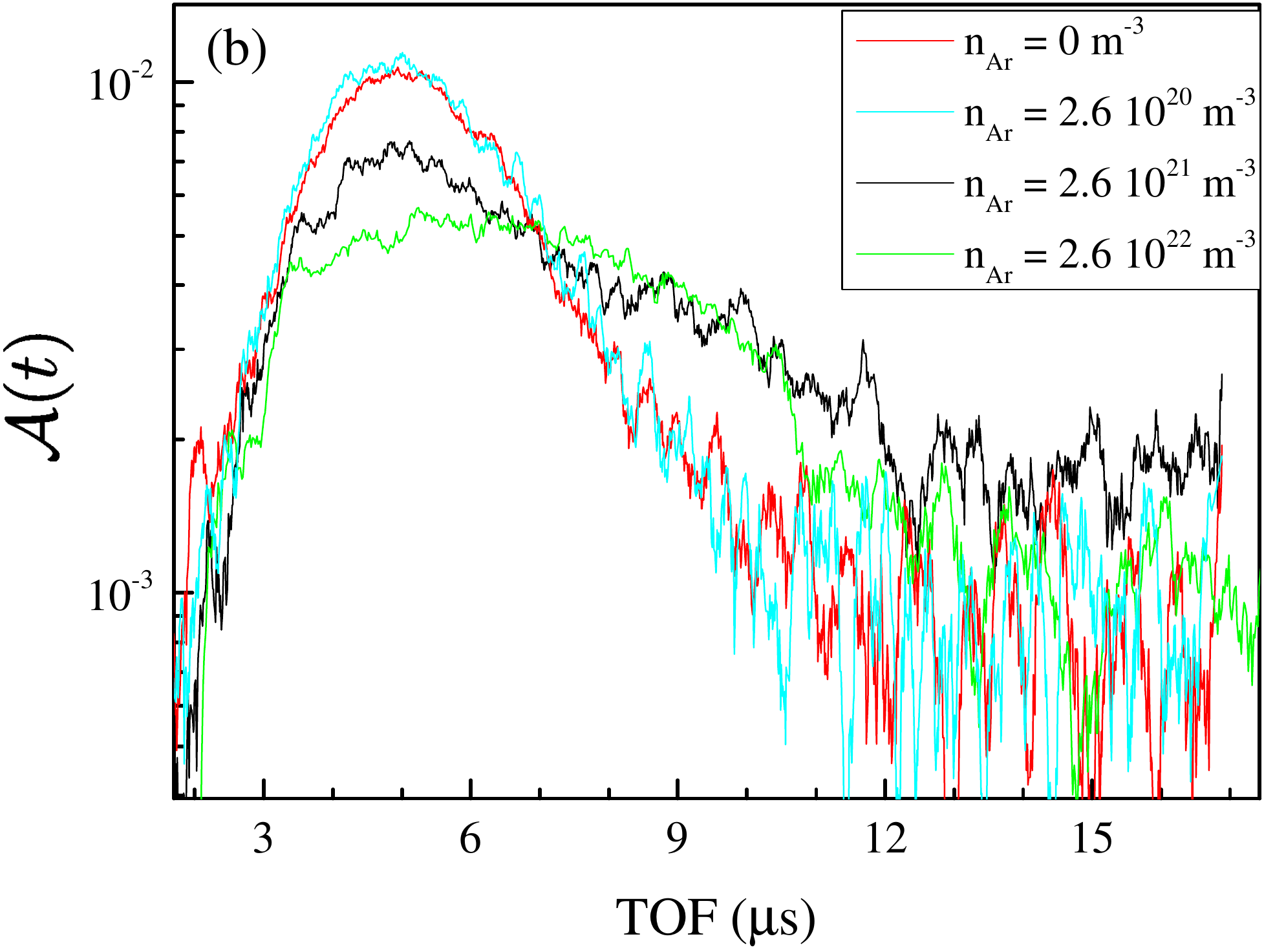}
	\caption{Translational cooling and diffusion of ablated BaH by buffer gas.  (a) Spectral widths of the absorption resonance have been measured at different TOFs in the presence of He, showing a good agreement with exponential decay.  The inset demonstrates the change of the absorption spectrum without the buffer gas (red squares) and after a 1.1 $\mu$s TOF (blue cicles).  (b) The transition from ballistic to diffusive expansion of BaH molecules in an Ar buffer gas.}
	\label{fig:HeBuff1}	
	\end{figure}
The translational temperatures can be determined by recording Doppler-broadened spectra for BaH molecules.  Figure \ref{fig:HeBuff1}(a) shows typical Doppler temperatures measured at different time intervals from the ablation pulse in the presence of room-temperature He buffer gas.  The inset features two spectra, one without the buffer gas and corresponding to the initial temperature, and another one after a 1.1 $\mu$s TOF.  The temperature follows an exponential decay, as predicted by Eq. (\ref{eq:buffcooling}), for both He and Ar buffer gasses.  The fitted thermalization rates are used to estimate the elastic cross sections $\sigma_{\mathrm{BaH},\mathrm{He}}=(0.31\pm0.02)\times10^{-20}$ m$^2$ and $\sigma_{\mathrm{BaH},\mathrm{Ar}}= (8\pm3)\times10^{-20}$ m$^{2}$.  The result for Ar is similar to the previously measured $\sigma_{\mathrm{YbF},\mathrm{He}}$, which has been found to underestimate the actual cross section by an order of magnitude \cite{skoff}.

Another method to estimate the cross section consists of measuring the diffusion time constants from time-resolved absorption. Figure~\ref{fig:HeBuff1}(b) shows the evolution of BaH absorption 3.5 mm from the target with the Ar buffer gas pressure.  By fitting the tails of the absorption profiles with an exponential decay we estimate $\sigma_{\mathrm{BaH},\mathrm{Ar}}^{\mathrm{(D)}}\approx(19\pm9)\times10^{-20}$ m$^2$.  It is clear from the profiles that for times longer than the typical ballistic arrival time ($\sim10$ $\mu$s) the data is dominated by technical noise, which suggests that $\sigma_{\mathrm{BaH},\mathrm{Ar}}^{\mathrm{(D)}}$ should be considered a lower bound (further suggesting that $\sigma_{\mathrm{BaH},\mathrm{bg}}$ were underestimated via the Doppler method).  For comparison, we repeated the diffusion measurement with ablated Yb atoms.  The resulting cross section is $\sigma_{\mathrm{Yb},\mathrm{Ar}}^{\mathrm{(D)}}= (1.9\pm0.1)\times10^{-17}$ m$^2$, improving the previously reported upper limit for Yb-Ar elastic collisions~\cite{yodh}.

The BaH-He elastic cross section, $\sigma_{\mathrm{BaH},\mathrm{He}}$, is a crucial parameter for determining the brightness of a He buffer gas cooled beam of BaH. The beam extraction efficiency of a molecule from the buffer has cell can be characterized by the dimensionless ratio, $\gamma_{\mathrm{cell}}$, between the time it takes a diffusing molecule to stick to the cell wall and the He exit time from the cell \cite{DoyleHutzlerCR12_BufferGasBeams}.  The brightest beams (up to 50\% extraction efficiency) are achieved with $\gamma_{\mathrm{cell}}>1$ \cite{Patterson}.  This ratio is directly proportional to $\sigma_{\mathrm{BaH},\mathrm{He}}$ and the buffer gas density, therefore the comparatively low value of $\sigma_{\mathrm{BaH},\mathrm{He}}$ can be compensated by cell geometry (such as the volume or target positioning) and experimental parameters that optimize the buffer gas density.

\subsubsection{Rotational thermalization}
\label{sec:RotationalCooling}

\begin{figure}
	\centering
	\subfigure[]{
	\includegraphics[width=0.48\textwidth]{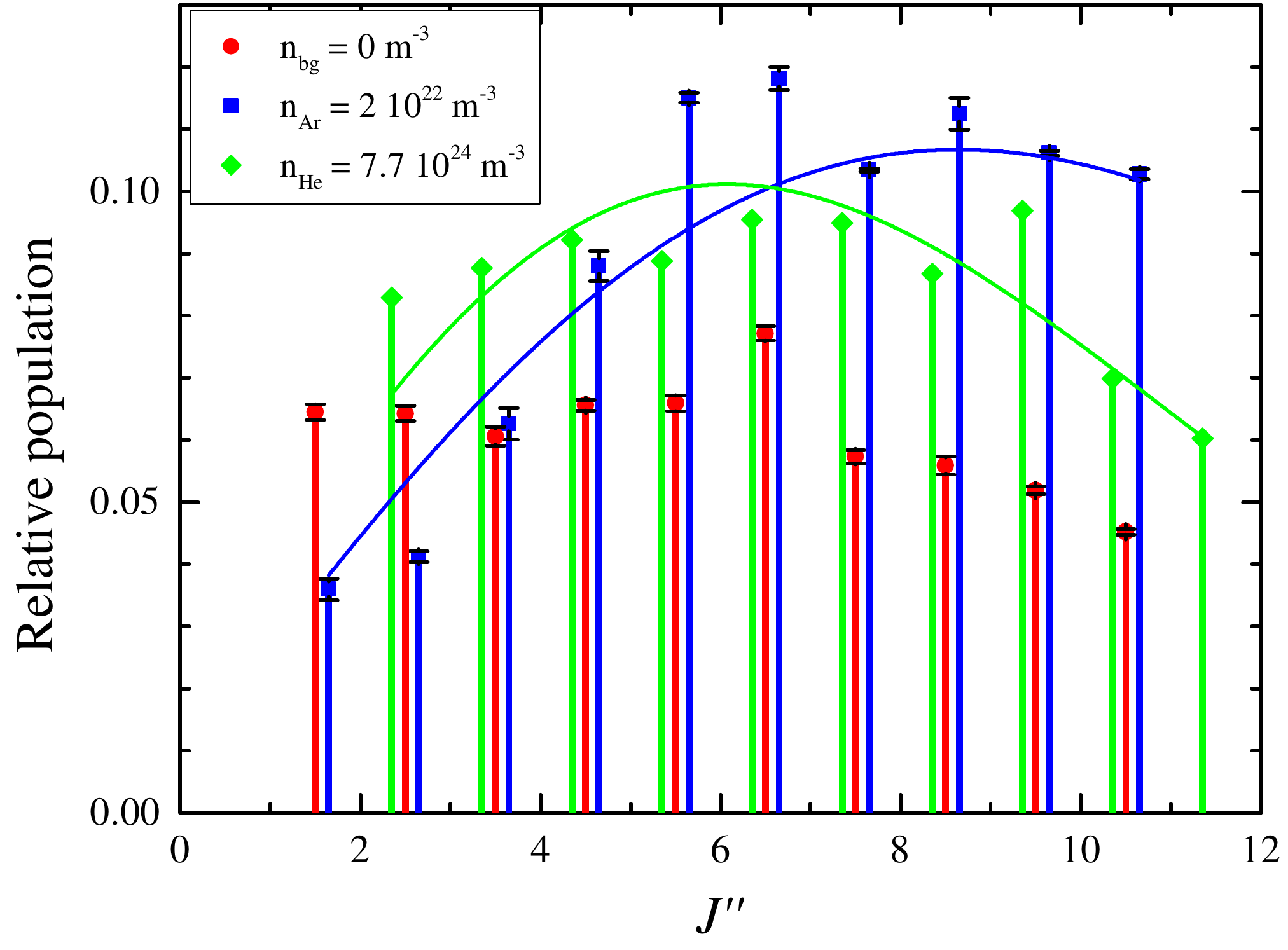}}
	\subfigure[]{
	\includegraphics[width=0.48\textwidth]{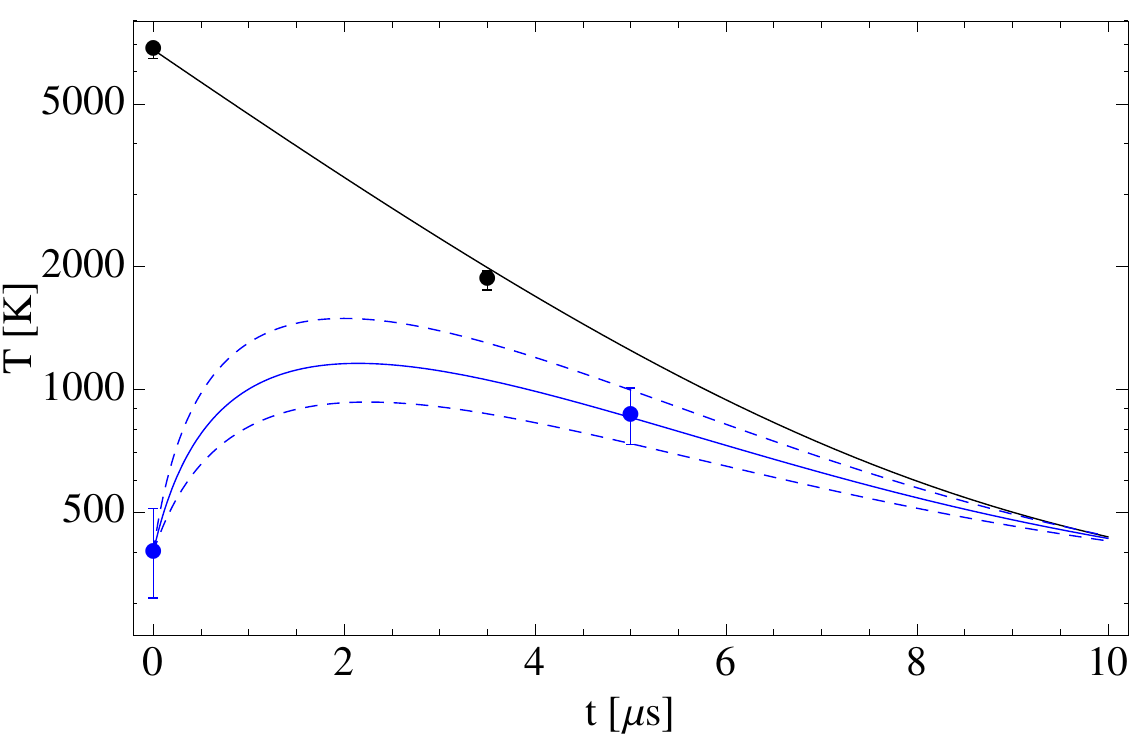}}
	\caption{Rotational cooling of ablated BaH by Ar and He buffer gases.  (a) $P_2$ absorption lines for various ground-state rotational level, for determining the rotational temperatures with and without a room temperature buffer gas.  (b) The measured translational and vibrational thermalization for room temperature Ar at the density of $2\times10^{22}$ m$^{-3}$, along with the fitted translational thermalization curve (black solid line) and the rotational thermalization curve with a thermalization rate chosen to best describe the data (blue solid line, with dashed lines showing the range consistent with the data).}
	\label{fig:HeBuff2}	
	\end{figure}
In addition to translational cooling, we studied the evolution of an initial non-thermal rotational distribution of BaH, described in Sec.~\ref{sec:ablation}, in the presence of a buffer gas.  The rotational temperature $T_r$ is estimated by fitting a Boltzmann distribution to the relative strengths of the $P_2$ absorption lines originating from the lowest eleven ground-state rotational levels ($J=1.5$ to $11.5$), as shown in Fig.~\ref{fig:HeBuff2}(a).

Collisions of BaH with both Ar and He buffer gases at room temperature lead to absorption enhancement at higher $J$ and a better agreement with a thermalized Maxwell-Boltzmann distribution.  For Ar buffer gas, $T_r=860\pm140$ K at $p=0.79$ torr after 5 $\mu$s TOF, while for He, $T_r=480\pm40$ K at $p=40$ torr after 2 $\mu$s TOF.  This result is in agreement with a thermalization process modeled by an exponential decay with the equilibrium temperature set by the instantaneous translational temperature.  In the case of Ar that is shown in Fig. \ref{fig:HeBuff2}(b), the rotational thermalization rate is about $7\pm2.5$ $\mu$s -- about twice the estimated translational thermalization rate -- in agreement with previous results that report rotational cross sections smaller than or similar to the corresponding elastic cross section values~\cite{DeMilleBarryPCCP11_CryogenicMolecularBeams,skoff}.

\subsection{Linear and nonlinear rovibrational spectroscopy of the $\mathrm{B}-\mathrm{X}\;(0,0)$ band}
\label{sec:LinNon}

We have performed precise rotational spectroscopy of the $\mathrm{B}^2\Sigma^+(v'=0)\leftarrow\mathrm{X}^2\Sigma^+(v"=0)$ electronic transition by measuring the absorption lines of the two $P$ branches ($\Delta J = \Delta N = -1$) and the satellite $^PQ_{12}$ branch ($\Delta J = 0,\,\Delta N = -1$).  The identification of the observed resonances was based on the lines reported in Ref.~\cite{watson} and on the values expected from the equilibrium constants~\cite{Appel85}.  The calculated resonance positions assumed that both the X and B states have pure Hund's case (b) couplings.  Hence their rotational energies $F_{1,2}(N)$ are split into two sub-manifolds via the spin-rotation interaction \cite{Brown},
\begin{eqnarray}
\nonumber F_{1,2}(N)=& BN(N+1)-DN^2(N+1)^2 \\
+& HN^3(N+1)^3\pm\frac{\gamma_N}{2}(N+\frac{1}{2}\mp\frac{1}{2}),
\end{eqnarray}
where the subscripts refer to $J=N\pm1/2$, and the effect of centrifugal distortion
is approximated as $\gamma_N = \gamma+\gamma_D N(N+1)$.

\begingroup
\begin{table}
\caption{\label{tab:Values} The measured transition lines and a comparison with previous data, for $P_1$, $P_2$, and $^PQ_{12}$ branches. The uncertainties represent a combination of statistical and systematic error. The last column summarizes the differences between our high-resolution measurements and Ref.~\cite{Appel85}.}
\begin{ruledtabular}
\begin{tabular}{l c c c c c l l}
 & $J''$ & This work & Prior work \cite{watson} & Theory~\cite{Appel85}& Difference \\
&& (nm) & (nm) & (nm) & (pm)\\
\hline
$P_1$ &1.5 & 905.31970(3) & -- & 905.32024 & -0.54(3)\\
& 2.5 & 906.09010(8)&906.0991& 906.09132& -1.22(8)\\
& 3.5 & 906.88122(4) & 906.8896 & 906.88223 & -1.01(4)\\
& 4.5 & 907.69219(3) & 907.6947 & 907.69279 & -0.60(3)\\
& 5.5 & 908.52230(4) & 908.5375 & 908.52281& -0.51(4)\\
& 6.5 & 909.37130(3) & 909.3769 & 909.37212 & -0.82(3)\\
& 7.5 & 910.23967(7) & 910.2295 & 910.24055 &-0.88(7)\\
& 8.5 & 911.12713(6) & 911.1335 & 911.12796&-0.83(6)\\
& 9.5 & 912.03306(3) & 912.0310 & 912.03418&-1.12(3)\\
& 10.5 & 912.95755(3) & 912.9627 & 912.95912&-1.57(3)\\
& 11.5 & 913.90056(3) & 912.9627 & 913.90260&-2.04(3)\\
\hline
$P_2$& 1.5 & 905.46645(3) & -- & 905.46699 & -0.54(3)\\
& 2.5 & 905.85052(6) & 905.8521 & 905.85112 & -0.60(6)\\
& 3.5 & 906.25389(2) & 906.2593 & 906.25420 & -0.31(2)\\
& 4.5 & 906.67616(4) & 906.6816 & 906.67614 & 0.02(4)\\
& 5.5 & 907.11708(4) & 907.1307 & 907.11690 & 0.18(4)\\
& 6.5 & 907.57678(3) & 907.5851 & 907.57642 & 0.36(3)\\
& 7.5 & 908.05521(3) & 908.0656 & 908.05467 & 0.54(3)\\
& 8.5 & 908.55110(2) & 908.5590 & 908.55162 & -0.52(2)\\
& 9.5 & 909.06738(2) & 909.0777 & 909.06724 & 0.14(2)\\
& 10.5 & 909.60222(4) & 909.6168& 909.60153& 0.69(4)\\
& 11.5 & 910.15485(4) & 910.1599 & 910.15449 & 0.36(4)\\
\hline
$^PQ_{12}$ & 0.5 & 905.29604(2)& -- & 905.29664& -0.60(2)\\
& 1.5 & 906.05214(3) & -- & 906.05193 & 0.21(3)\\
& 2.5 & 906.82640(8) & -- & 906.82702 & -0.62(8)\\
& 3.5 & 907.62107(7) & -- & 907.62171 & -0.64(7)\\
\end{tabular}
\end{ruledtabular}
\end{table}
\endgroup
\begin{figure}
	\centering
	\includegraphics[width=0.48\textwidth]{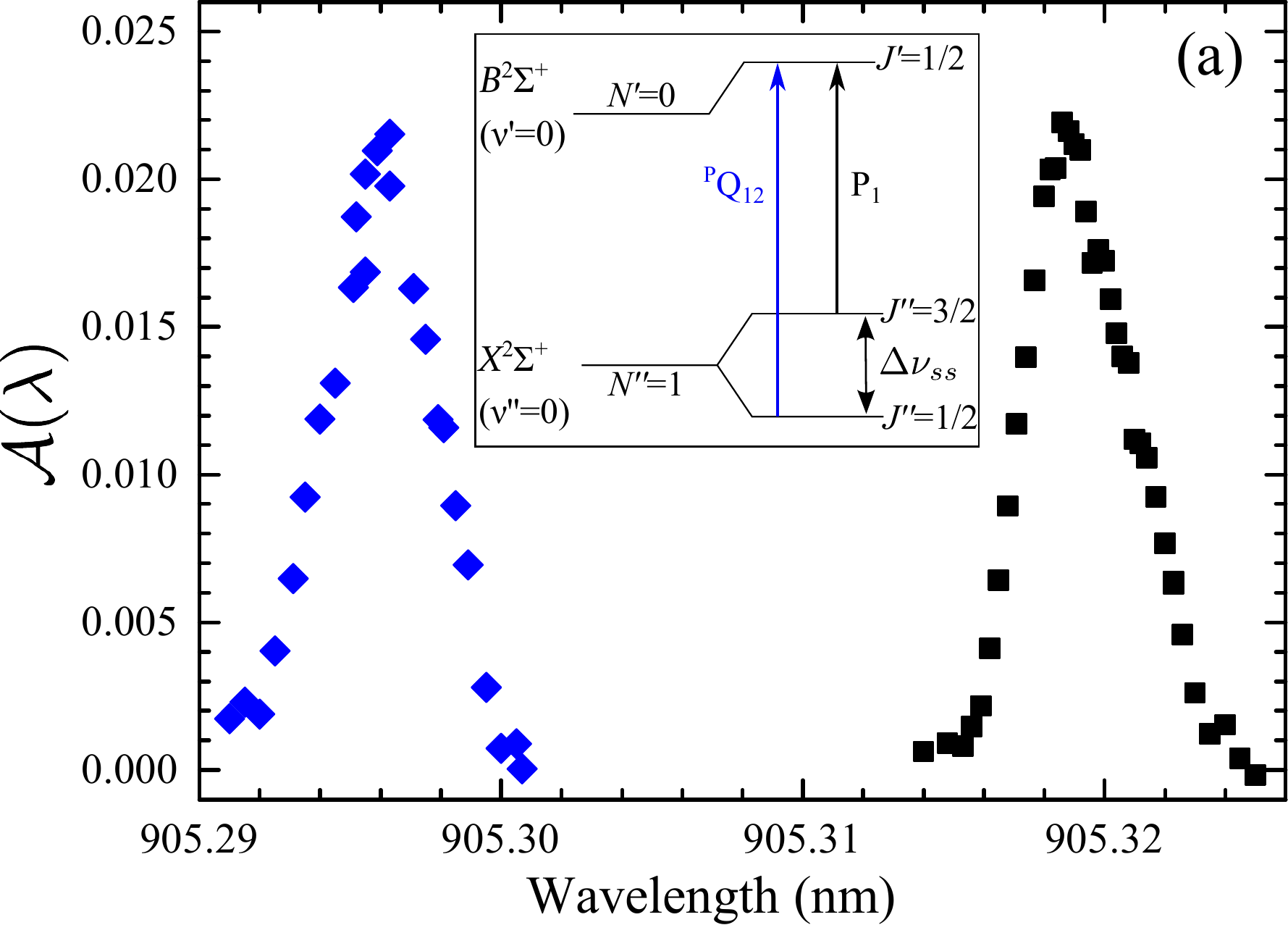}
	\includegraphics[width=0.48\textwidth]{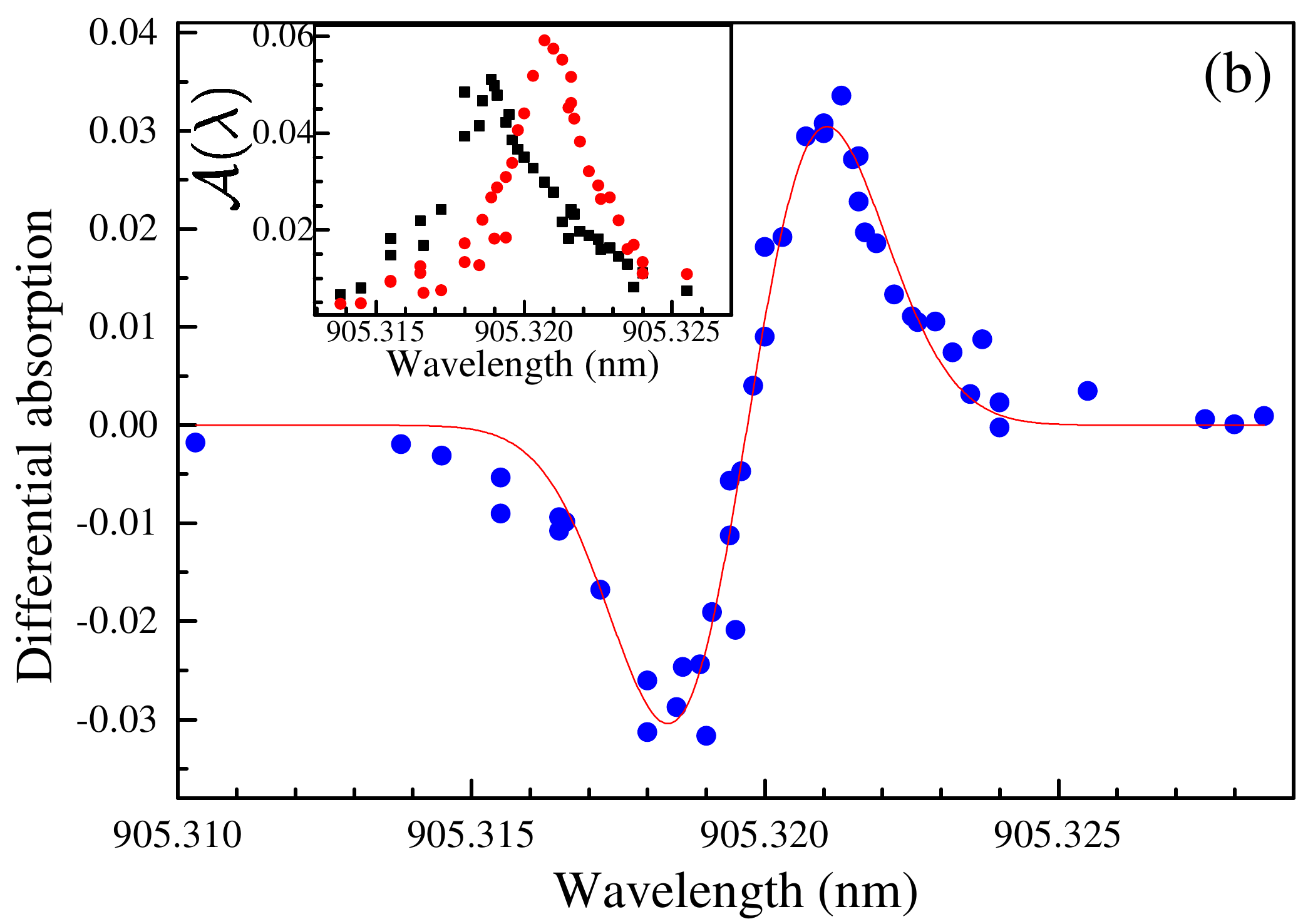}
	\caption{$\mathrm{B}^2\Sigma^+(N=0)\mathrm{X}^2\Sigma^+(N=1)$ spectroscopy.  (a) Determination of the $\mathrm{X}^2\Sigma^+(N=1)$ level spin-rotation splitting. (b) Doppler shift measurement for transverse spectroscopy on the $P_1(1.5)$ cooling transition.}
	\label{fig:fff}
	\end{figure}
By tuning the mirror angle and position in the cavity of our ECDL, we spanned the rotational spectrum of the $\mathrm{B}-\mathrm{X}\;(0,0)$ band between 905 and 911 nm. We identified all the $P_1$ and $P_2$ branch lines from $J=1.5$ to $11.5$, and $^PQ_{12}$ lines from $J=0.5$ to $3.5$.  The transition wavelengths are listed in Table ~\ref{tab:Values}.  The measured values deviate from the expected wavelengths by about $-180$ MHz on average with a standard deviation of 150 MHz,
which is consistent with the 200 MHz accuracy of the wavemeter.  The sample spectral lineshapes in Fig.~\ref{fig:fff}(a) exhibit typical Gaussian widths of $\sim3$ pm.

The ground state spin splitting $\Delta\nu_{ss}$ is of particular interest for laser cooling of BaH.  It has been shown that rotational branching is minimized by cooling on the $N'=0\leftarrow N''=1$ transition \cite{stuhl}.
For the excited level $\mathrm{B}^2\Sigma^+(N'=0,J'=0.5)$, parity selection rules allow only two possible decay channels (neglecting hyperfine structure):  a $P_1$ transition to $J = 1.5$, and a satellite $^PQ_{12}$ transition to $J=0.5$ with the same rotational line strength \cite{Mulliken}.  The measured spectrum of this spin-rotation doublet is shown in Fig. \ref{fig:fff}(a).  The well-resolved line shapes yield the spin-rotation splitting $\Delta\nu_{ss}(N=1)$ = 8.654(3) GHz centered at about 905.3 nm.  For the purposes of laser cooling, this splitting can be spanned with existing electro-optical technology.

Particularly precise measurements of both the $P_1(1.5)$ and the $^PQ_{12}(0.5)$ transition wavelengths have been performed via Doppler-free interrogation.  The Doppler shift is the dominant systematic effect on ablated BaH spectroscopy.  This is due to the non-zero angle between the probe laser propagation and the direction perpendicular to the plume forward velocity, which depends on the positioning of each BaH$_2$ sample and its surface regularity.  We assessed the Doppler shift by aligning a pair of counterpropagating probe beams with mrad precision and probing the ablation plume either simultaneously or consecutively, as shown in Fig. \ref{fig:fff}(b).  The differential absorption signal from each photodetector is centered on the Doppler-free resonant wavelength, while common noise sources such as the ablation yield and laser frequency variations cancel out, increasing the signal-to-noise ratio and leading to an uncertainty of only 3.6 MHz.

\begin{figure}
\begin{center}
\includegraphics[width=.48\textwidth]{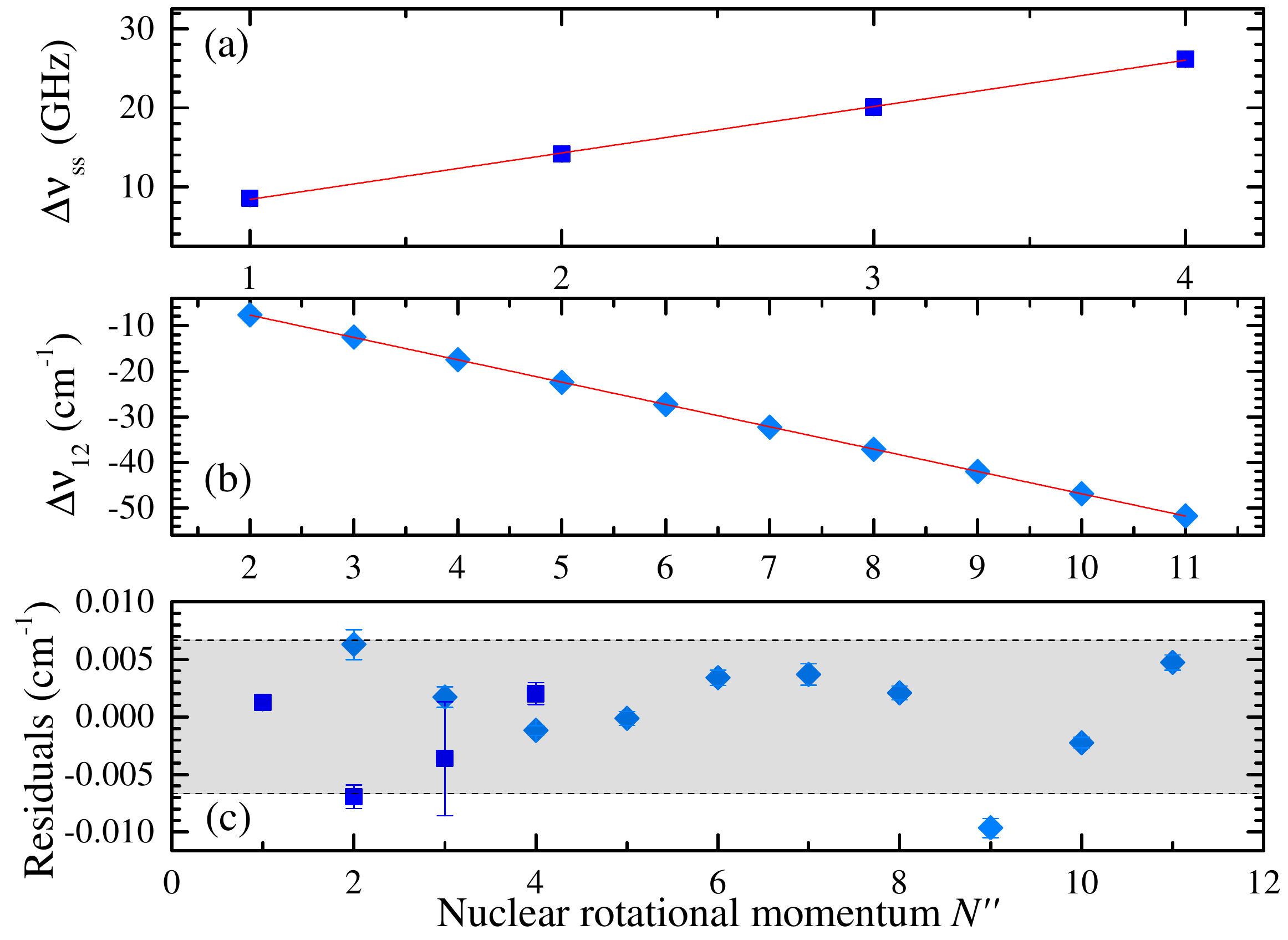}
\caption{Ground-state (a) spin splitting and (b) $P$-branch line splitting measurement with low-$N$ levels of the $\mathrm{B}-\mathrm{X}\;(0,0)$ band.  (c) Combined fit residuals.  The gray area represents the 200 MHz wavemeter uncertainty.}
\label{fig:LS&SS}
\end{center}
\end{figure}
It is possible to improve the accuracy of the ground state $\Delta\nu_{ss}$ by measuring the frequency difference between each $P_1$ line and its satellite transition $^PQ_{12}$ as a function of $N$ (which depends only on the spin-rotation constant $\gamma_N$) and the line splitting of the two $P$ branches,
\beq
\Delta\nu_{12}(P) \equiv P_1(N)-P_2(N) = (\gamma_B-\gamma_X)N-\frac{1}{2}(\gamma_B+\gamma_X),
\eeq
where the $\gamma$'s are the spin-rotation constants for the B and X states.  Both the line splitting and the spin splitting data (10 points and 4 points, respectively) were used to determine the spin-rotation constants, as shown in Fig. \ref{fig:LS&SS}.  The resulting values are $\gamma_X=0.19165(88)$ cm$^{-1}$, $\gamma_B=-4.7532(12)$ cm$^{-1}$, and $\gamma_{BD}=0.0003297(55)$ cm$^{-1}$.
The ground state centrifugal distortion constant was neglected because it is expected to be on the order of $10^{-5}$ cm$^{-1}$. The agreement with Ref. \cite{Appel85}, where the number of recorded transitions is in the hundreds, is at the fourth significant digit for both linear spin-rotation constants.

\subsection{Direct measurement of the Franck-Condon factors to the lowest two ground state vibrational bands}
\label{sec:FCFMeas}

\begin{figure}
	\centering
	\includegraphics[width=0.486\textwidth]{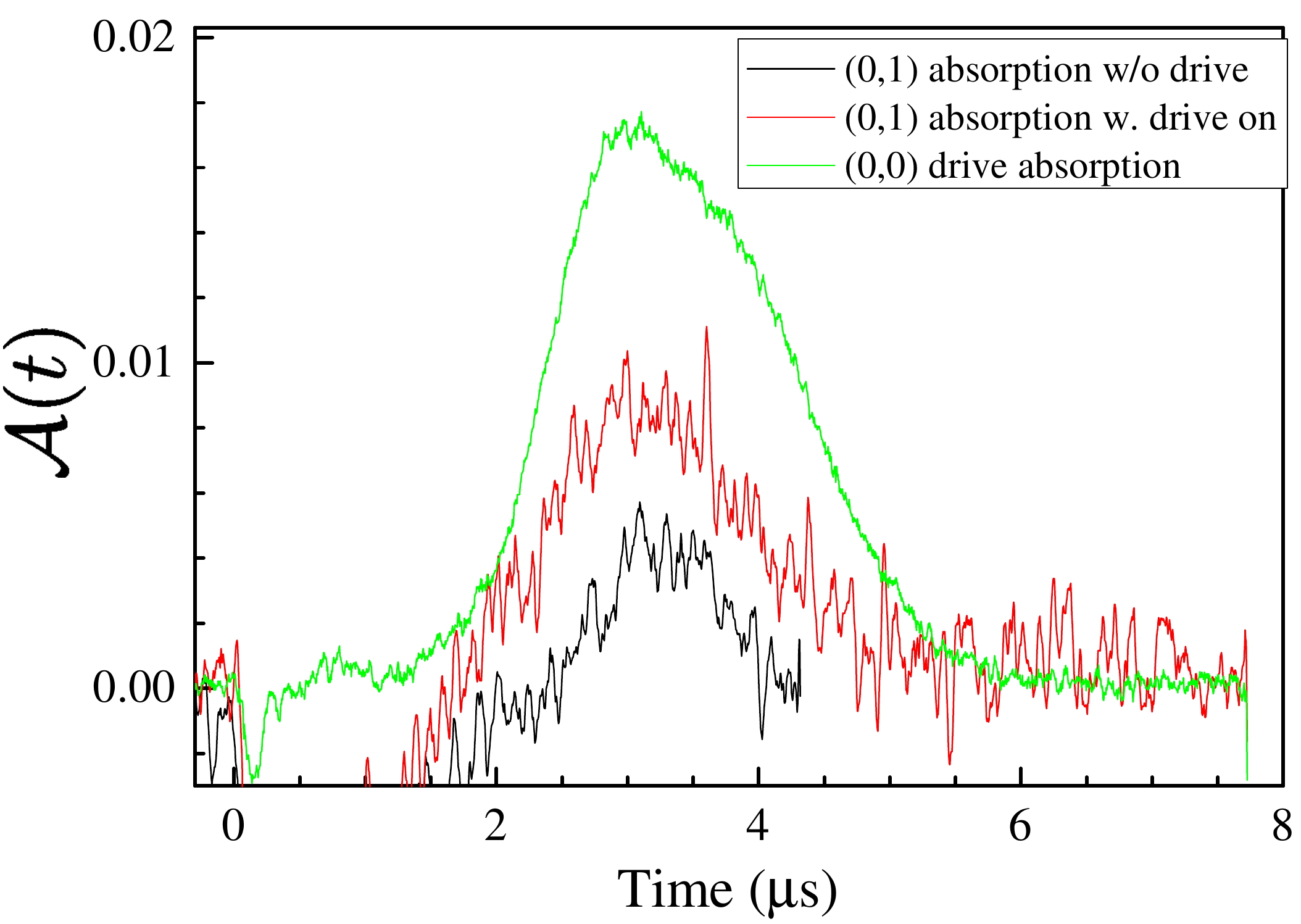}
	\caption{Dual resonant absorption spectroscopy of the $\mathrm{B}-\mathrm{X}\;(0,0)$ and $\mathrm{B}-\mathrm{X}\;(0,1)$ transitions [$P_1(1.5)$].}  
	\label{fig:FCmeas}
\end{figure}
We measured the decay branching ratio of the $B(v'=0)$ energy level to the first two ground state vibrational levels $v'' =0,1$ by dual resonant absorption spectroscopy.  A laser beam from an additional ECDL tuned to $\lambda_{01}=1009.43$ nm was superimposed onto the 905 nm laser to simultaneously probe the $|\mathrm{B},v'=0,N'=0,J'=0.5\rangle-|\mathrm{X},v''=0,N''=0,J''=0.5\rangle$ and $|\mathrm{B},0,0,0.5\rangle\leftarrow|\mathrm{X},1,0,0.5\rangle$ $P_1$ transitions.  A weak resonant absorption signal was observed on the latter transition, both with and without the 905 nm pump laser, as shown in Fig.~\ref{fig:FCmeas}.

\begin{figure}
	\centering
	\includegraphics[width=0.486\textwidth]{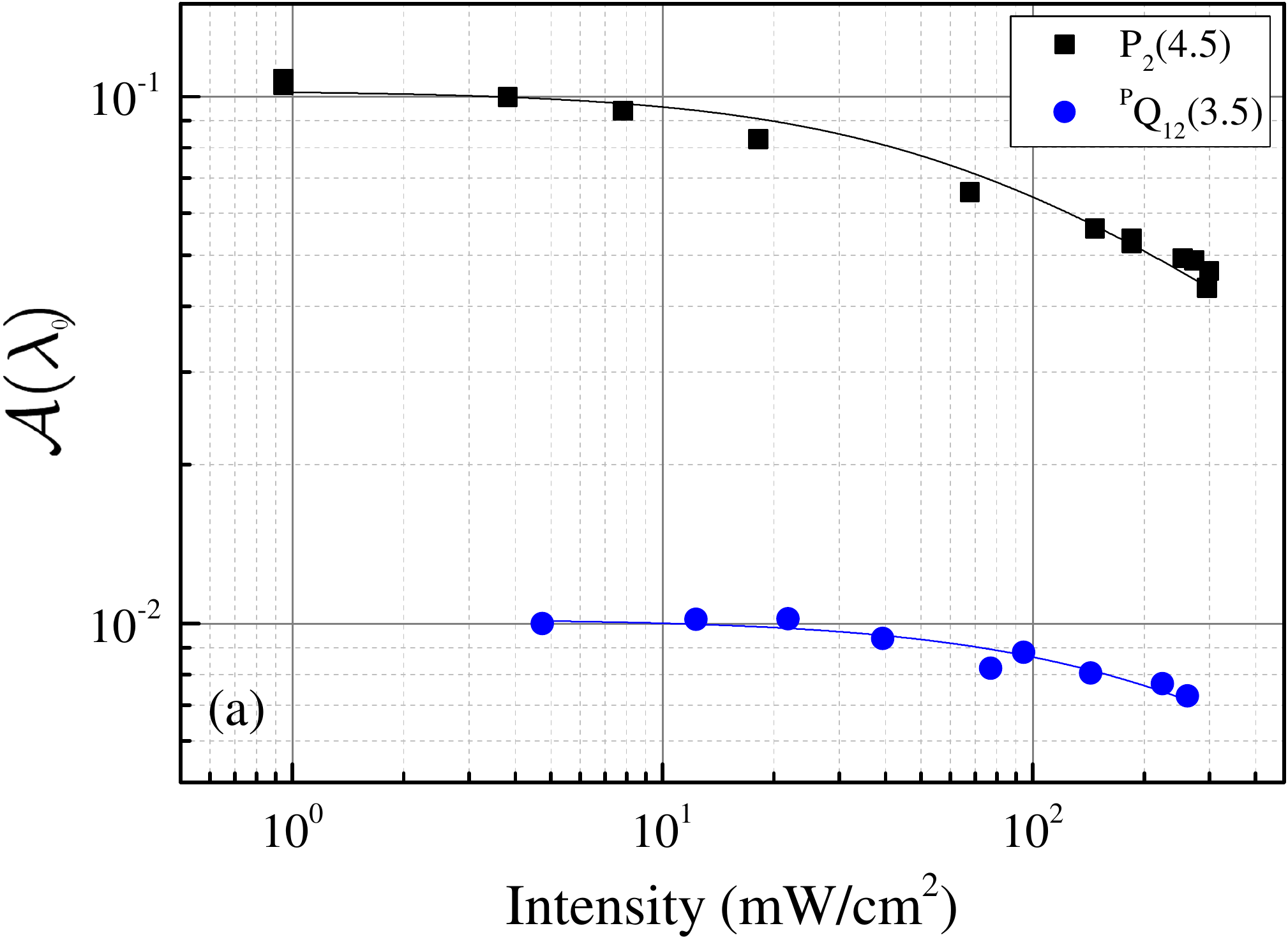}
	\includegraphics[width=0.486\textwidth]{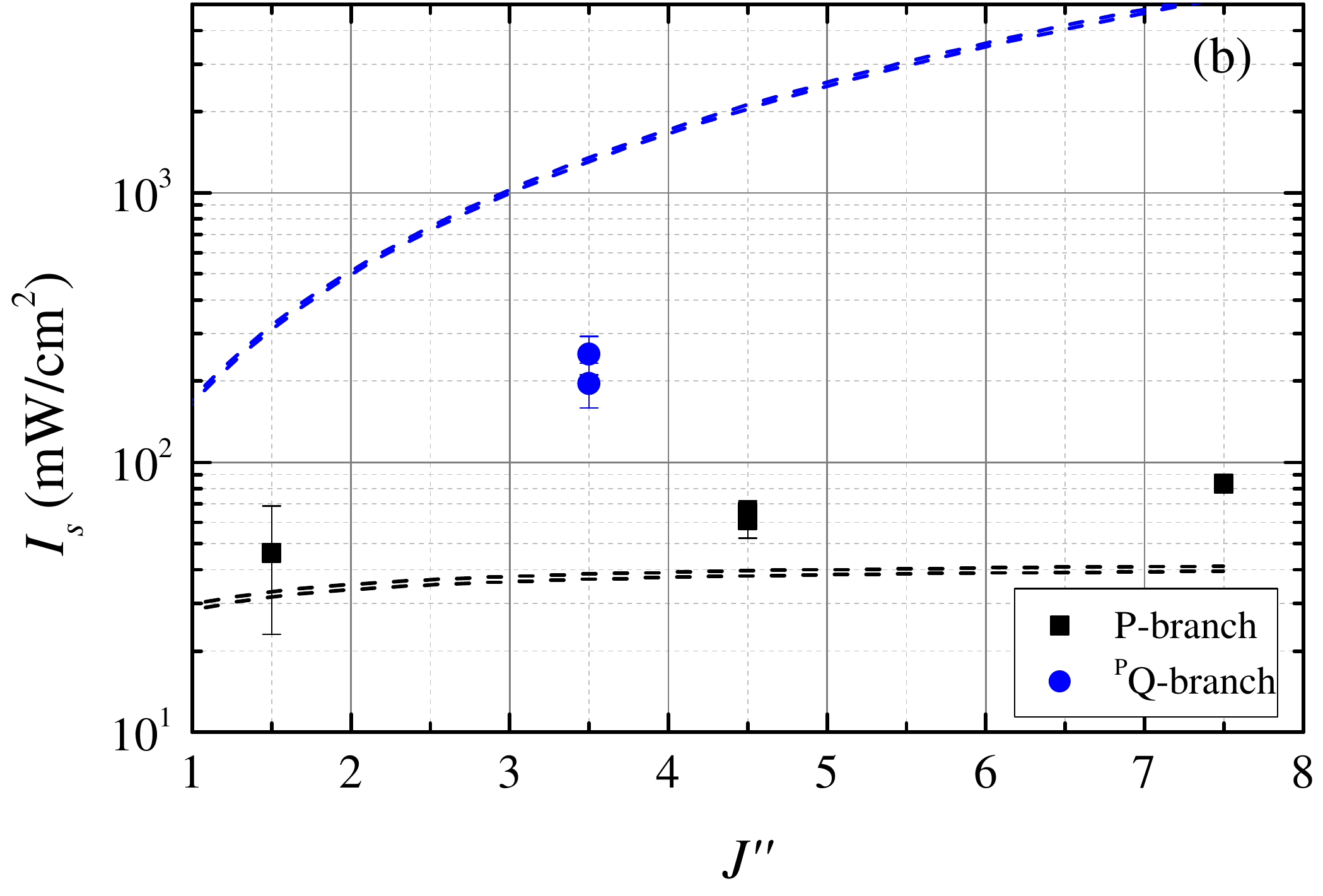}
	\caption{Experimental study of the transient saturation intensity on a BaH molecular plume for different rovibrational transitions. (a) The fractional absorption versus probe laser intensity for the $P_2(J=4.5)$ and $^PQ_{12}(3.5)$ lines. (b) Summary of all the $I_s$ measurements on the $P$ (black squares) and $^PQ$ (blue circles) branches of the $\mathrm{B}-\mathrm{X}\;(0,0)$ band.  Comparison lines with the expected values according to Eq. (\ref{eq:Isat}) are drawn.}
	\label{fig:sat}
\end{figure}

We used Eq.~\ref{eq:Adiff} to estimate the decay branching ratio, assuming that $R_{00}+R_{01}\approx1$ \cite{LanePRA15_HFromBaH,rama}.  The saturation parameter $s_0$ was experimentally determined by varying the pump power and recording $\mathcal{A}_{00}^0$, as shown in Fig.~\ref{fig:sat} (a) for two sample transitions.  For this particular measurement we used the value $\mathcal{A}_{00}^0 = 0.038(10)$, where the uncertainty represents the large absorption fluctuations.  Assuming a thermal vibrational distribution of the molecules ($f_1 = 0.44$ at $T= 7000$ K), the resulting decay branching ratio to the first excited vibrational level is
\beq
R_{01} = 0.043_{-0.030}^{+0.018}.
\label{eq:R01Result}
\eeq
This estimate is in good agreement with \emph{ab-initio} calculations \cite{LanePRA15_HFromBaH}, and can be also expressed in terms of the Franck-Condon factor $q_{01}=0.059_{-0.041}^{+0.025}$.

An alternative method of the branching ratio measurement is via the study of saturation intensity $I_s(N'',J'')$, as described in Sec.~\ref{sec:sat}. We performed resonant absorption spectroscopy on several $P$ and $^PQ$ lines by varying the probe laser intensity and measured the fractional absorption, as shown in Fig.\ref{fig:sat} (a).  The data is fitted with the function $\mathcal{A}=\mathcal{A}_0/\sqrt{1+x/I_s(N'',J'')}$ \cite{skoff}.

While the saturation behavior of the absorption signal follows the expected trend, the fitted $I_s$ values disagree with Eqs. (\ref{eq:Isat},\ref{eq:R01Result}), as shown in Fig.~\ref{fig:sat}(b).  For the $P$-branch, the growth with $J''$ is larger than predicted by Eq. (\ref{eq:Isat}), which could be explained with a dependence of the total radiative decay rate $\Gamma$ on the rotational number $J''$~\cite{berg}. We believe that the disagreement of the $^PQ$ saturation intensity values is partly due to the limited range of probe intensities used in our measurements.

\section{Outlook}
\label{sec:outlook}

The ablation and spectroscopy described here provide the groundwork needed to create a bright cryogenic molecular beam of BaH, and to laser cool and eventually trap BaH into a molecular magneto-optical trap (MOT).
Helium buffer gas cooling provides a proven starting point for ultracold molecule production process \cite{Campbell}, and our work can be extended to a cryogenic environment where laser-ablated BaH molecules would thermalize with a flowing He buffer gas at 4 K.  In particular, Sections \ref{sec:ablation} and \ref{sec:buffer} demonstrate the amenability of BaH to extraction via laser ablation of a solid target and to buffer-gas rotational and translational cooling, both of which are key components of the future cryogenic beam.  Cryogenic buffer gas cooling combined with expansion cooling is expected to lead to molecular beam temperatures of around 1-4 K, with high ground-state molecule densities of $\sim10^9/$cm$^3$ and low forward velocities of 10's m/s \cite{Patterson}.

\begingroup
\begin{table}
\caption{\label{gfactors}The $g$-factors relevant to laser cooling BaH on the $\mathrm{B}^2\Sigma^+\leftarrow\mathrm{X}^2\Sigma^+$ transition, with and without $J$-mixing between hyperfine states.}
\begin{ruledtabular}
\begin{tabular}{l c c}
State & Ideal case & With state mixing\\
\hline
$X^2\Sigma^+(N=1, J=3/2, F=2)$ & 0.500 & 0.500 \\
$X^2\Sigma^+(N=1, J=3/2, F=1)$ & 0.833 & 0.836 \\
$X^2\Sigma^+(N=1, J=1/2, F=1)$ & -0.333 & -0.336 \\
$X^2\Sigma^+(N=1, J=1/2, F=0)$ & 0.000 & 0.000 \\
\hline
$B^2\Sigma^+(N=0, J=1/2, F=1)$ & 1.000 & 0.938\\
$B^2\Sigma^+(N=0, J=1/2, F=0)$ & 0.000 & 0.000\\
\end{tabular}
\end{ruledtabular}
\end{table}
\endgroup
Figure \ref{fig:lasercooling} illustrates two possible laser cooling transitions, $\mathrm{A}^2\Pi\leftarrow\mathrm{X}^2\Sigma^+$ and $\mathrm{B}^2\Sigma^+\leftarrow\mathrm{X}^2\Sigma^+$, focusing on the latter in part (b) since it
is expected to be more favorable for creating a MOT.  The magnetic $g$-factors of the upper and lower states are comparable in this case, which is a determining factor for the strength of the trapping forces \cite{tarbMOT}.
Each $g$-factor must account for the spin-rotation and hyperfine interactions.  The effective molecular $g_F$ with respect to the total angular momentum $\mathbf{F} = \mathbf{J}+\mathbf{I}$  is
$g_F \approx  g_J[F(F+1) + J(J+1) - I(I+1)]/[2F(F+1)]$,
where $g_J=g_s\,[J(J+1) + S(S+1) - N(N+1)]/[2J(J+1)]$ is the effective spin-rotation $g$-factor for a Hund's case (b) $^2\Sigma$ state, and $g_s$ = 2. In the expression for $g_F$ we have neglected the term of order $\mu_N/\mu_B$ arising from the nuclear magnetic moment, as well as the mixing between different $J$ states which is valid at low external fields.  To include the mixing, we solved the Hamiltonian exactly, and the relevant results are listed in Tab. \ref{gfactors}.  For the $\mathrm{B}^2\Sigma^+(N=0, J=1/2)$ state, $g_J=1$, in contrast to the $\mathrm{A}^2\Pi_{1/2}$ excited state where $g_J\approx0$.
Therefore, the MOT forces are minimized for any light polarization configuration in the case of $\mathrm{A}^2\Pi_{1/2}-\mathrm{X}^2\Sigma^+$ cooling \cite{tarbMOT}, whereas for $\mathrm{B}^2\Sigma-\mathrm{X}^2\Sigma^+$ cooling we predict a $\sim5$-fold enhancement of the typical molecular optical trapping forces \cite{Barry,DeMilleImprovedMOT} without the need for radio-frequency remixing of dark states \cite{hummon}.

BaH has a low lying $\mathrm{A}'^2\Delta_{3/2}$ state (Fig. \ref{fig:lasercooling}) that can provide a loss channel for molecules cycled through the $\mathrm{A}^2\Pi$ or $\mathrm{B}^2\Sigma^+$ states \cite{yeo2015}.  While cooling via $\mathrm{B}^2\Sigma^+$ reduces this possibility, it is important to verify that spontaneous decay to the lower lying electronic states does not significantly limit the number of molecule-photon scattering events.  The B, A, and A$'$ states form a single complex \cite{bernard}, and therefore electronically forbidden $\Sigma-\Delta$ transitions may become weakly allowed.  Spontaneous decay to $\mathrm{A}^2\Pi$ (branching ratio $R_{\Sigma-\Pi}$ in Fig. \ref{fig:lasercooling}) takes place at the long wavelength of $\sim6$ $\mu$m and is therefore suppressed by the wavelength factor of $\sim3\times10^{-3}$.  In addition, most molecules lost to $\mathrm{A}^2\Pi$ are quickly returned to the cooling cycle.
The transition moment from $\mathrm{B}^2\Sigma^+$ to $\mathrm{A}'^2\Delta_{3/2}$ is $\sim2\%$ of that to the ground state due to a $1\%$ mixing of A and A$'$ \cite{barrow}.  Accounting for the additional wavelength suppression factor, we expect the loss to be sufficiently low.  However, the molecules that decay to the A$'$ state are not returned to the cooling cycle, and this loss channel has to be further quantified .

We experimentally confirmed the large vibrational branching ratio $R_{00}\sim0.96$, which is favorable for establishing a nearly cycling laser cooling transition. Direct measurements of branching into the higher vibrational levels can be made with the cryogenic molecular beam using laser induced fluorescence \cite{TarbuttZhuangPCCP11_FCFsYbF}.

In conclusion, we present precise absorption spectroscopy on an ablation plume of BaH to demonstrate the practicality of laser ablation of these molecules, to investigate their thermalization with buffer gases, and to provide new and improved measurements of the wavelengths and molecular constants needed for operating a BaH cryogenic beam source and for laser cooling.  Hyperfine and Zeeman structure will be resolved with the help of the cryogenic molecular beam.

\begin{acknowledgments}
We thank F. Apfelbeck, A. T. Grier, F. S\"{o}rensen, and K. Verteletsky for their contributions to this work.  We acknowledge partial support by the ONR grant N000-14-14-1-0802, and G.Z.I. acknowledges support by the NSF IGERT Grant DGE-1069240.
\end{acknowledgments}


\begin{thebibliography}{49}
\expandafter\ifx\csname natexlab\endcsname\relax\def\natexlab#1{#1}\fi
\expandafter\ifx\csname bibnamefont\endcsname\relax
  \def\bibnamefont#1{#1}\fi
\expandafter\ifx\csname bibfnamefont\endcsname\relax
  \def\bibfnamefont#1{#1}\fi
\expandafter\ifx\csname citenamefont\endcsname\relax
  \def\citenamefont#1{#1}\fi
\expandafter\ifx\csname url\endcsname\relax
  \def\url#1{\texttt{#1}}\fi
\expandafter\ifx\csname urlprefix\endcsname\relax\def\urlprefix{URL }\fi
\providecommand{\bibinfo}[2]{#2}
\providecommand{\eprint}[2][]{\url{#2}}

\bibitem[{\citenamefont{Herzberg}(1950)}]{Herzberg}
\bibinfo{author}{\bibfnamefont{G.}~\bibnamefont{Herzberg}},
  \emph{\bibinfo{title}{Spectra of Diatomic Molecules}},
  vol.~\bibinfo{volume}{I} of \emph{\bibinfo{series}{Molecular Spectra and
  Molecular Structure}} (\bibinfo{publisher}{Van Nostrand},
  \bibinfo{address}{New York}, \bibinfo{year}{1950}).

\bibitem[{\citenamefont{Di~Rosa}(2004)}]{DiRosa}
\bibinfo{author}{\bibfnamefont{M.~D.} \bibnamefont{Di~Rosa}},
  \bibinfo{journal}{Eur. Phys. J. D} \textbf{\bibinfo{volume}{31}},
  \bibinfo{pages}{395} (\bibinfo{year}{2004}).

\bibitem[{\citenamefont{Carr et~al.}(2009)\citenamefont{Carr, DeMille, Krems,
  and Ye}}]{Carr}
\bibinfo{author}{\bibfnamefont{L.~D.} \bibnamefont{Carr}},
  \bibinfo{author}{\bibfnamefont{D.}~\bibnamefont{DeMille}},
  \bibinfo{author}{\bibfnamefont{R.~V.} \bibnamefont{Krems}}, \bibnamefont{and}
  \bibinfo{author}{\bibfnamefont{J.}~\bibnamefont{Ye}}, \bibinfo{journal}{New.
  J. Phys} \textbf{\bibinfo{volume}{11}}, \bibinfo{pages}{055049}
  (\bibinfo{year}{2009}).

\bibitem[{\citenamefont{Lane}(2015)}]{LanePRA15_HFromBaH}
\bibinfo{author}{\bibfnamefont{I.~C.} \bibnamefont{Lane}},
  \bibinfo{journal}{Phys. Rev. A} \textbf{\bibinfo{volume}{92}},
  \bibinfo{pages}{022511} (\bibinfo{year}{2015}).

\bibitem[{\citenamefont{Fredrickson and Watson}(1932)}]{fredwatson}
\bibinfo{author}{\bibfnamefont{W.}~\bibnamefont{Fredrickson}} \bibnamefont{and}
  \bibinfo{author}{\bibfnamefont{W.}~\bibnamefont{Watson}},
  \bibinfo{journal}{Phys. Rev.} \textbf{\bibinfo{volume}{39}},
  \bibinfo{pages}{753} (\bibinfo{year}{1932}).

\bibitem[{\citenamefont{Watson}(1933)}]{watson}
\bibinfo{author}{\bibfnamefont{W.}~\bibnamefont{Watson}},
  \bibinfo{journal}{Phys. Rev.} \textbf{\bibinfo{volume}{43}},
  \bibinfo{pages}{9} (\bibinfo{year}{1933}).

\bibitem[{\citenamefont{Kopp et~al.}(1966)\citenamefont{Kopp, Kronekvist, and
  Guntsch}}]{kopp}
\bibinfo{author}{\bibfnamefont{I.}~\bibnamefont{Kopp}},
  \bibinfo{author}{\bibfnamefont{M.}~\bibnamefont{Kronekvist}},
  \bibnamefont{and} \bibinfo{author}{\bibfnamefont{A.}~\bibnamefont{Guntsch}},
  \bibinfo{journal}{Ark. Fysik.} \textbf{\bibinfo{volume}{32}},
  \bibinfo{pages}{371} (\bibinfo{year}{1966}).

\bibitem[{\citenamefont{Appelblad et~al.}(1985)\citenamefont{Appelblad, Berg,
  and Klynning}}]{Appel85}
\bibinfo{author}{\bibfnamefont{O.}~\bibnamefont{Appelblad}},
  \bibinfo{author}{\bibfnamefont{L.~E.} \bibnamefont{Berg}}, \bibnamefont{and}
  \bibinfo{author}{\bibfnamefont{L.}~\bibnamefont{Klynning}},
  \bibinfo{journal}{Phys. Scr.} \textbf{\bibinfo{volume}{31}},
  \bibinfo{pages}{69} (\bibinfo{year}{1985}).

\bibitem[{\citenamefont{Barrow et~al.}(1991)\citenamefont{Barrow, Howard,
  Bernard, and Effantin}}]{barrow}
\bibinfo{author}{\bibfnamefont{R.}~\bibnamefont{Barrow}},
  \bibinfo{author}{\bibfnamefont{B.~J.} \bibnamefont{Howard}},
  \bibinfo{author}{\bibfnamefont{A.}~\bibnamefont{Bernard}}, \bibnamefont{and}
  \bibinfo{author}{\bibfnamefont{C.}~\bibnamefont{Effantin}},
  \bibinfo{journal}{Mol. Phys.} \textbf{\bibinfo{volume}{72}},
  \bibinfo{pages}{971} (\bibinfo{year}{1991}).

\bibitem[{\citenamefont{Ram and Bernath}(2013)}]{Bernath}
\bibinfo{author}{\bibfnamefont{R.}~\bibnamefont{Ram}} \bibnamefont{and}
  \bibinfo{author}{\bibfnamefont{P.}~\bibnamefont{Bernath}},
  \bibinfo{journal}{J. Mol. Spectrosc.} \textbf{\bibinfo{volume}{283}},
  \bibinfo{pages}{18} (\bibinfo{year}{2013}).

\bibitem[{\citenamefont{Ramanaiah and Lakshman}(1982)}]{rama}
\bibinfo{author}{\bibfnamefont{M.~V.} \bibnamefont{Ramanaiah}}
  \bibnamefont{and} \bibinfo{author}{\bibfnamefont{S.~V.~J.}
  \bibnamefont{Lakshman}}, \bibinfo{journal}{Physica B+C}
  \textbf{\bibinfo{volume}{113}}, \bibinfo{pages}{263} (\bibinfo{year}{1982}).

\bibitem[{\citenamefont{Tarbutt}(2015)}]{tarbMOT}
\bibinfo{author}{\bibfnamefont{M.~R.} \bibnamefont{Tarbutt}},
  \bibinfo{journal}{New. J. Phys} \textbf{\bibinfo{volume}{17}},
  \bibinfo{pages}{015007} (\bibinfo{year}{2015}).

\bibitem[{\citenamefont{Shuman et~al.}(2010)\citenamefont{Shuman, Barry, and
  DeMille}}]{shuman}
\bibinfo{author}{\bibfnamefont{E.~S.} \bibnamefont{Shuman}},
  \bibinfo{author}{\bibfnamefont{J.~F.} \bibnamefont{Barry}}, \bibnamefont{and}
  \bibinfo{author}{\bibfnamefont{D.}~\bibnamefont{DeMille}},
  \bibinfo{journal}{Nature} \textbf{\bibinfo{volume}{467}},
  \bibinfo{pages}{820} (\bibinfo{year}{2010}).

\bibitem[{\citenamefont{Hummon et~al.}(2013)\citenamefont{Hummon, Yeo, Stuhl,
  Collopy, Xia, and Ye}}]{hummon}
\bibinfo{author}{\bibfnamefont{M.~T.} \bibnamefont{Hummon}},
  \bibinfo{author}{\bibfnamefont{M.}~\bibnamefont{Yeo}},
  \bibinfo{author}{\bibfnamefont{B.~K.} \bibnamefont{Stuhl}},
  \bibinfo{author}{\bibfnamefont{A.~L.} \bibnamefont{Collopy}},
  \bibinfo{author}{\bibfnamefont{Y.}~\bibnamefont{Xia}}, \bibnamefont{and}
  \bibinfo{author}{\bibfnamefont{J.}~\bibnamefont{Ye}}, \bibinfo{journal}{Phys.
  Rev. Lett.} \textbf{\bibinfo{volume}{110}}, \bibinfo{pages}{143001}
  (\bibinfo{year}{2013}).

\bibitem[{\citenamefont{Zhelyazkova et~al.}(2014)\citenamefont{Zhelyazkova,
  Cournol, Wall, Matsushima, Hudson, Hinds, Tarbutt, and Sauer}}]{zhelya}
\bibinfo{author}{\bibfnamefont{V.}~\bibnamefont{Zhelyazkova}},
  \bibinfo{author}{\bibfnamefont{A.}~\bibnamefont{Cournol}},
  \bibinfo{author}{\bibfnamefont{T.~E.} \bibnamefont{Wall}},
  \bibinfo{author}{\bibfnamefont{A.}~\bibnamefont{Matsushima}},
  \bibinfo{author}{\bibfnamefont{J.~J.} \bibnamefont{Hudson}},
  \bibinfo{author}{\bibfnamefont{E.~A.} \bibnamefont{Hinds}},
  \bibinfo{author}{\bibfnamefont{M.~R.} \bibnamefont{Tarbutt}},
  \bibnamefont{and} \bibinfo{author}{\bibfnamefont{B.~E.} \bibnamefont{Sauer}},
  \bibinfo{journal}{Phys. Rev. A} \textbf{\bibinfo{volume}{89}},
  \bibinfo{pages}{053416} (\bibinfo{year}{2014}).

\bibitem[{\citenamefont{Barry et~al.}(2011)\citenamefont{Barry, Shuman, and
  DeMille}}]{DeMilleBarryPCCP11_CryogenicMolecularBeams}
\bibinfo{author}{\bibfnamefont{J.~F.} \bibnamefont{Barry}},
  \bibinfo{author}{\bibfnamefont{E.~S.} \bibnamefont{Shuman}},
  \bibnamefont{and} \bibinfo{author}{\bibfnamefont{D.}~\bibnamefont{DeMille}},
  \bibinfo{journal}{Phys. Chem. Chem. Phys.} \textbf{\bibinfo{volume}{13}},
  \bibinfo{pages}{18936} (\bibinfo{year}{2011}).

\bibitem[{\citenamefont{Hutzler et~al.}(2012)\citenamefont{Hutzler, Lu, and
  Doyle}}]{DoyleHutzlerCR12_BufferGasBeams}
\bibinfo{author}{\bibfnamefont{N.~R.} \bibnamefont{Hutzler}},
  \bibinfo{author}{\bibfnamefont{H.-I.} \bibnamefont{Lu}}, \bibnamefont{and}
  \bibinfo{author}{\bibfnamefont{J.~M.} \bibnamefont{Doyle}},
  \bibinfo{journal}{Chem. Rev.} \textbf{\bibinfo{volume}{112}},
  \bibinfo{pages}{4803} (\bibinfo{year}{2012}).

\bibitem[{\citenamefont{Barry et~al.}(2014)\citenamefont{Barry, McCarron,
  Norrgard, Steinecker, and DeMille}}]{Barry}
\bibinfo{author}{\bibfnamefont{J.~F.} \bibnamefont{Barry}},
  \bibinfo{author}{\bibfnamefont{D.~J.} \bibnamefont{McCarron}},
  \bibinfo{author}{\bibfnamefont{E.~N.} \bibnamefont{Norrgard}},
  \bibinfo{author}{\bibfnamefont{M.~H.} \bibnamefont{Steinecker}},
  \bibnamefont{and} \bibinfo{author}{\bibfnamefont{D.}~\bibnamefont{DeMille}},
  \bibinfo{journal}{Nature} \textbf{\bibinfo{volume}{512}},
  \bibinfo{pages}{286} (\bibinfo{year}{2014}).

\bibitem[{\citenamefont{Harnafi and Dubreuil}(1991)}]{harnafi}
\bibinfo{author}{\bibfnamefont{M.}~\bibnamefont{Harnafi}} \bibnamefont{and}
  \bibinfo{author}{\bibfnamefont{B.}~\bibnamefont{Dubreuil}},
  \bibinfo{journal}{J. Appl. Phys.} \textbf{\bibinfo{volume}{69}},
  \bibinfo{pages}{7565} (\bibinfo{year}{1991}).

\bibitem[{\citenamefont{Cheung et~al.}(1991)\citenamefont{Cheung, Ying, Zheng,
  and Kwok}}]{cheung}
\bibinfo{author}{\bibfnamefont{N.}~\bibnamefont{Cheung}},
  \bibinfo{author}{\bibfnamefont{Q.}~\bibnamefont{Ying}},
  \bibinfo{author}{\bibfnamefont{J.}~\bibnamefont{Zheng}}, \bibnamefont{and}
  \bibinfo{author}{\bibfnamefont{H.}~\bibnamefont{Kwok}}, \bibinfo{journal}{J.
  Appl. Phys.} \textbf{\bibinfo{volume}{69}}, \bibinfo{pages}{6349}
  (\bibinfo{year}{1991}).

\bibitem[{\citenamefont{Geohegan and Mashborn}(1989)}]{geohegan}
\bibinfo{author}{\bibfnamefont{D.~B.} \bibnamefont{Geohegan}} \bibnamefont{and}
  \bibinfo{author}{\bibfnamefont{D.~N.} \bibnamefont{Mashborn}},
  \bibinfo{journal}{Appl. Phys. Lett.} \textbf{\bibinfo{volume}{55}},
  \bibinfo{pages}{2345} (\bibinfo{year}{1989}).

\bibitem[{\citenamefont{Bushaw and Alexander}(1998)}]{bushaw}
\bibinfo{author}{\bibfnamefont{B.~A.} \bibnamefont{Bushaw}} \bibnamefont{and}
  \bibinfo{author}{\bibfnamefont{M.~L.} \bibnamefont{Alexander}},
  \bibinfo{journal}{Appl. Surf. Sci.} \textbf{\bibinfo{volume}{127--129}},
  \bibinfo{pages}{935} (\bibinfo{year}{1998}).

\bibitem[{\citenamefont{Eason}(2007)}]{PLD}
\bibinfo{editor}{\bibfnamefont{R.}~\bibnamefont{Eason}}, ed.,
  \emph{\bibinfo{title}{Pulsed laser deposition of thin films}}
  (\bibinfo{publisher}{Wiley}, \bibinfo{address}{Hoboken, New Jersey},
  \bibinfo{year}{2007}).

\bibitem[{\citenamefont{Belouet}(1996)}]{Belouet}
\bibinfo{author}{\bibfnamefont{C.}~\bibnamefont{Belouet}},
  \bibinfo{journal}{Appl. Surf. Sci.} \textbf{\bibinfo{volume}{96--98}},
  \bibinfo{pages}{630} (\bibinfo{year}{1996}).

\bibitem[{\citenamefont{Kools et~al.}(1992)\citenamefont{Kools, Baller,
  De~Zwart, and Dieleman}}]{Kools}
\bibinfo{author}{\bibfnamefont{J.~C.~S.} \bibnamefont{Kools}},
  \bibinfo{author}{\bibfnamefont{T.~S.} \bibnamefont{Baller}},
  \bibinfo{author}{\bibfnamefont{S.~T.} \bibnamefont{De~Zwart}},
  \bibnamefont{and} \bibinfo{author}{\bibfnamefont{J.}~\bibnamefont{Dieleman}},
  \bibinfo{journal}{J. Appl. Phys.} \textbf{\bibinfo{volume}{71}},
  \bibinfo{pages}{4547} (\bibinfo{year}{1992}).

\bibitem[{\citenamefont{Kelly and Dreyfus}(1988)}]{Kelly}
\bibinfo{author}{\bibfnamefont{R.}~\bibnamefont{Kelly}} \bibnamefont{and}
  \bibinfo{author}{\bibfnamefont{R.~W.} \bibnamefont{Dreyfus}},
  \bibinfo{journal}{Surf. Sci.} \textbf{\bibinfo{volume}{198}},
  \bibinfo{pages}{263} (\bibinfo{year}{1988}).

\bibitem[{\citenamefont{Gerginov and Tanner}(2003)}]{georginov}
\bibinfo{author}{\bibfnamefont{V.}~\bibnamefont{Gerginov}} \bibnamefont{and}
  \bibinfo{author}{\bibfnamefont{C.~E.} \bibnamefont{Tanner}},
  \bibinfo{journal}{Optics Communications} \textbf{\bibinfo{volume}{222}},
  \bibinfo{pages}{17} (\bibinfo{year}{2003}).

\bibitem[{\citenamefont{Wall et~al.}(2008)\citenamefont{Wall, Kanem, Hudson,
  Sauer, Cho, Boshier, Hinds, and Tarbutt}}]{Wall}
\bibinfo{author}{\bibfnamefont{T.~E.} \bibnamefont{Wall}},
  \bibinfo{author}{\bibfnamefont{J.~F.} \bibnamefont{Kanem}},
  \bibinfo{author}{\bibfnamefont{J.~J.} \bibnamefont{Hudson}},
  \bibinfo{author}{\bibfnamefont{B.~E.} \bibnamefont{Sauer}},
  \bibinfo{author}{\bibfnamefont{D.}~\bibnamefont{Cho}},
  \bibinfo{author}{\bibfnamefont{M.~G.} \bibnamefont{Boshier}},
  \bibinfo{author}{\bibfnamefont{E.~A.} \bibnamefont{Hinds}}, \bibnamefont{and}
  \bibinfo{author}{\bibfnamefont{M.~R.} \bibnamefont{Tarbutt}},
  \bibinfo{journal}{Phys. Rev. A} \textbf{\bibinfo{volume}{78}},
  \bibinfo{pages}{062509} (\bibinfo{year}{2008}).

\bibitem[{\citenamefont{Berg et~al.}(1997)\citenamefont{Berg, Ekvall,
  Hishikawa, and Kelly}}]{berg}
\bibinfo{author}{\bibfnamefont{L.~E.} \bibnamefont{Berg}},
  \bibinfo{author}{\bibfnamefont{K.}~\bibnamefont{Ekvall}},
  \bibinfo{author}{\bibfnamefont{A.}~\bibnamefont{Hishikawa}},
  \bibnamefont{and} \bibinfo{author}{\bibfnamefont{S.}~\bibnamefont{Kelly}},
  \bibinfo{journal}{Phys. Scr.} \textbf{\bibinfo{volume}{55}},
  \bibinfo{pages}{269} (\bibinfo{year}{1997}).

\bibitem[{\citenamefont{Watson}(2008)}]{Watson20085}
\bibinfo{author}{\bibfnamefont{J.~K.} \bibnamefont{Watson}},
  \bibinfo{journal}{J. Mol. Spectrosc.} \textbf{\bibinfo{volume}{252}},
  \bibinfo{pages}{5 } (\bibinfo{year}{2008}).

\bibitem[{\citenamefont{Demtr\"oder}(2008)}]{demtr}
\bibinfo{author}{\bibfnamefont{W.}~\bibnamefont{Demtr\"oder}},
  \emph{\bibinfo{title}{Laser Spectroscopy}}, vol. \bibinfo{volume}{2,
  Experimental techniques} (\bibinfo{publisher}{Springer},
  \bibinfo{year}{2008}), \bibinfo{edition}{4th} ed.

\bibitem[{\citenamefont{{American Elements, private
  communication}}()}]{AmericanElementsPrivate}
\bibinfo{author}{\bibnamefont{{American Elements, private communication}}}.

\bibitem[{\citenamefont{Lu et~al.}(2011)\citenamefont{Lu, Rasmussen, Wright,
  Patterson, and Doyle}}]{Lu}
\bibinfo{author}{\bibfnamefont{H.-I.} \bibnamefont{Lu}},
  \bibinfo{author}{\bibfnamefont{J.}~\bibnamefont{Rasmussen}},
  \bibinfo{author}{\bibfnamefont{M.}~\bibnamefont{Wright}},
  \bibinfo{author}{\bibfnamefont{D.}~\bibnamefont{Patterson}},
  \bibnamefont{and} \bibinfo{author}{\bibfnamefont{J.~M.} \bibnamefont{Doyle}},
  \bibinfo{journal}{Phys. Chem. Chem. Phys.} \textbf{\bibinfo{volume}{13}},
  \bibinfo{pages}{18986} (\bibinfo{year}{2011}).

\bibitem[{\citenamefont{Magg et~al.}(1988)\citenamefont{Magg, Birk, and
  Jones}}]{Magg}
\bibinfo{author}{\bibfnamefont{U.}~\bibnamefont{Magg}},
  \bibinfo{author}{\bibfnamefont{H.}~\bibnamefont{Birk}}, \bibnamefont{and}
  \bibinfo{author}{\bibfnamefont{H.}~\bibnamefont{Jones}},
  \bibinfo{journal}{Chem. Phys. Lett.} \textbf{\bibinfo{volume}{149}},
  \bibinfo{pages}{321} (\bibinfo{year}{1988}).

\bibitem[{\citenamefont{Dreyfus et~al.}(1986)\citenamefont{Dreyfus, Kelly, and
  Walkup}}]{rottemp}
\bibinfo{author}{\bibfnamefont{R.~W.} \bibnamefont{Dreyfus}},
  \bibinfo{author}{\bibfnamefont{R.}~\bibnamefont{Kelly}}, \bibnamefont{and}
  \bibinfo{author}{\bibfnamefont{R.~E.} \bibnamefont{Walkup}},
  \bibinfo{journal}{Appl. Phys. Lett.} \textbf{\bibinfo{volume}{49}},
  \bibinfo{pages}{1478} (\bibinfo{year}{1986}).

\bibitem[{\citenamefont{Campbell and Doyle}(2009)}]{Campbell}
\bibinfo{author}{\bibfnamefont{W.~C.} \bibnamefont{Campbell}} \bibnamefont{and}
  \bibinfo{author}{\bibfnamefont{J.~M.} \bibnamefont{Doyle}},
  \emph{\bibinfo{title}{Cold Molecules: theory, experiment, applications}}
  (\bibinfo{publisher}{CRC Press}, \bibinfo{year}{2009}), chap.
  \bibinfo{chapter}{Cooling, Trap Loading, and Beam Production Using a
  Cryogenic Helium Buffer Gas}, pp. \bibinfo{pages}{473--508}.

\bibitem[{\citenamefont{Weinstein et~al.}(1998)\citenamefont{Weinstein,
  {deCarvalho}, Guillet, Friedrich, and Doyle}}]{DoyleWeinsteinNature98_mKCaH}
\bibinfo{author}{\bibfnamefont{J.~D.} \bibnamefont{Weinstein}},
  \bibinfo{author}{\bibfnamefont{R.}~\bibnamefont{{deCarvalho}}},
  \bibinfo{author}{\bibfnamefont{T.}~\bibnamefont{Guillet}},
  \bibinfo{author}{\bibfnamefont{B.}~\bibnamefont{Friedrich}},
  \bibnamefont{and} \bibinfo{author}{\bibfnamefont{J.~M.} \bibnamefont{Doyle}},
  \bibinfo{journal}{Nature} \textbf{\bibinfo{volume}{395}},
  \bibinfo{pages}{148} (\bibinfo{year}{1998}).

\bibitem[{\citenamefont{Hutzler et~al.}(2011)\citenamefont{Hutzler, Parsons,
  Gurevich, Hess, Petrik, Spaun, Vutha, DeMille, Gabrielse, and
  Doyle}}]{Hutzler}
\bibinfo{author}{\bibfnamefont{N.~R.} \bibnamefont{Hutzler}},
  \bibinfo{author}{\bibfnamefont{M.~F.} \bibnamefont{Parsons}},
  \bibinfo{author}{\bibfnamefont{Y.~V.} \bibnamefont{Gurevich}},
  \bibinfo{author}{\bibfnamefont{P.~W.} \bibnamefont{Hess}},
  \bibinfo{author}{\bibfnamefont{E.}~\bibnamefont{Petrik}},
  \bibinfo{author}{\bibfnamefont{B.}~\bibnamefont{Spaun}},
  \bibinfo{author}{\bibfnamefont{A.~C.} \bibnamefont{Vutha}},
  \bibinfo{author}{\bibfnamefont{D.}~\bibnamefont{DeMille}},
  \bibinfo{author}{\bibfnamefont{G.}~\bibnamefont{Gabrielse}},
  \bibnamefont{and} \bibinfo{author}{\bibfnamefont{J.~M.} \bibnamefont{Doyle}},
  \bibinfo{journal}{Phys. Chem. Chem. Phys.} \textbf{\bibinfo{volume}{13}},
  \bibinfo{pages}{18976} (\bibinfo{year}{2011}).

\bibitem[{\citenamefont{Skoff et~al.}(2011)\citenamefont{Skoff, Hendricks,
  Sinclair, Hudson, Segal, Sauer, Hinds, and Tarbutt}}]{skoff}
\bibinfo{author}{\bibfnamefont{S.~M.} \bibnamefont{Skoff}},
  \bibinfo{author}{\bibfnamefont{R.~J.} \bibnamefont{Hendricks}},
  \bibinfo{author}{\bibfnamefont{C.~D.~J.} \bibnamefont{Sinclair}},
  \bibinfo{author}{\bibfnamefont{J.~J.} \bibnamefont{Hudson}},
  \bibinfo{author}{\bibfnamefont{D.~M.} \bibnamefont{Segal}},
  \bibinfo{author}{\bibfnamefont{B.~E.} \bibnamefont{Sauer}},
  \bibinfo{author}{\bibfnamefont{E.~A.} \bibnamefont{Hinds}}, \bibnamefont{and}
  \bibinfo{author}{\bibfnamefont{M.~R.} \bibnamefont{Tarbutt}},
  \bibinfo{journal}{Phys. Rev. A} \textbf{\bibinfo{volume}{83}},
  \bibinfo{pages}{023418} (\bibinfo{year}{2011}).

\bibitem[{\citenamefont{{R. deCarvalho} et~al.}(1999)\citenamefont{{R.
  deCarvalho}, Doyle, Friedrich, Guillet, Kim, Patterson, and
  Weinstein}}]{deCarv}
\bibinfo{author}{\bibnamefont{{R. deCarvalho}}},
  \bibinfo{author}{\bibfnamefont{J.~M.} \bibnamefont{Doyle}},
  \bibinfo{author}{\bibfnamefont{B.}~\bibnamefont{Friedrich}},
  \bibinfo{author}{\bibfnamefont{T.}~\bibnamefont{Guillet}},
  \bibinfo{author}{\bibfnamefont{J.}~\bibnamefont{Kim}},
  \bibinfo{author}{\bibfnamefont{D.}~\bibnamefont{Patterson}},
  \bibnamefont{and} \bibinfo{author}{\bibfnamefont{J.~D.}
  \bibnamefont{Weinstein}}, \bibinfo{journal}{Eur. Phys. J. D}
  \textbf{\bibinfo{volume}{7}}, \bibinfo{pages}{289} (\bibinfo{year}{1999}).

\bibitem[{\citenamefont{Yodh et~al.}(1985)\citenamefont{Yodh, Golub, and
  Mossberg}}]{yodh}
\bibinfo{author}{\bibfnamefont{A.~G.} \bibnamefont{Yodh}},
  \bibinfo{author}{\bibfnamefont{J.}~\bibnamefont{Golub}}, \bibnamefont{and}
  \bibinfo{author}{\bibfnamefont{T.~W.} \bibnamefont{Mossberg}},
  \bibinfo{journal}{Phys. Rev. A} \textbf{\bibinfo{volume}{32}},
  \bibinfo{pages}{844} (\bibinfo{year}{1985}).

\bibitem[{\citenamefont{Patterson and Doyle}(2007)}]{Patterson}
\bibinfo{author}{\bibfnamefont{D.}~\bibnamefont{Patterson}} \bibnamefont{and}
  \bibinfo{author}{\bibfnamefont{J.~M.} \bibnamefont{Doyle}},
  \bibinfo{journal}{J. Chem. Phys.} \textbf{\bibinfo{volume}{126}},
  \bibinfo{pages}{154307} (\bibinfo{year}{2007}).

\bibitem[{\citenamefont{Brown and Carrington}(2003)}]{Brown}
\bibinfo{author}{\bibfnamefont{J.~M.} \bibnamefont{Brown}} \bibnamefont{and}
  \bibinfo{author}{\bibfnamefont{A.}~\bibnamefont{Carrington}},
  \emph{\bibinfo{title}{Rotational Spectroscopy of Diatomic Molecules}}
  (\bibinfo{publisher}{Cambridge University Press}, \bibinfo{year}{2003}).

\bibitem[{\citenamefont{Stuhl et~al.}(2008)\citenamefont{Stuhl, Sawyer, Wang,
  and Ye}}]{stuhl}
\bibinfo{author}{\bibfnamefont{B.~K.} \bibnamefont{Stuhl}},
  \bibinfo{author}{\bibfnamefont{B.~C.} \bibnamefont{Sawyer}},
  \bibinfo{author}{\bibfnamefont{D.}~\bibnamefont{Wang}}, \bibnamefont{and}
  \bibinfo{author}{\bibfnamefont{J.}~\bibnamefont{Ye}}, \bibinfo{journal}{Phys.
  Rev. Lett.} \textbf{\bibinfo{volume}{101}}, \bibinfo{pages}{243002}
  (\bibinfo{year}{2008}).

\bibitem[{\citenamefont{Mulliken}(1931)}]{Mulliken}
\bibinfo{author}{\bibfnamefont{R.~S.} \bibnamefont{Mulliken}},
  \bibinfo{journal}{Rev. Mod. Phys.} \textbf{\bibinfo{volume}{3}},
  \bibinfo{pages}{89} (\bibinfo{year}{1931}).

\bibitem[{\citenamefont{McCarron et~al.}(2015)\citenamefont{McCarron, Norrgard,
  Steinecker, and DeMille}}]{DeMilleImprovedMOT}
\bibinfo{author}{\bibfnamefont{D.~J.} \bibnamefont{McCarron}},
  \bibinfo{author}{\bibfnamefont{E.~B.} \bibnamefont{Norrgard}},
  \bibinfo{author}{\bibfnamefont{M.~H.} \bibnamefont{Steinecker}},
  \bibnamefont{and} \bibinfo{author}{\bibfnamefont{D.}~\bibnamefont{DeMille}},
  \bibinfo{journal}{New. J. Phys.} \textbf{\bibinfo{volume}{17}},
  \bibinfo{pages}{035014} (\bibinfo{year}{2015}).

\bibitem[{\citenamefont{Yeo et~al.}(2015)\citenamefont{Yeo, Hummon, Collopy,
  Yan, Hemmerling, Chae, Doyle, and Ye}}]{yeo2015}
\bibinfo{author}{\bibfnamefont{M.}~\bibnamefont{Yeo}},
  \bibinfo{author}{\bibfnamefont{M.~T.} \bibnamefont{Hummon}},
  \bibinfo{author}{\bibfnamefont{A.~L.} \bibnamefont{Collopy}},
  \bibinfo{author}{\bibfnamefont{B.}~\bibnamefont{Yan}},
  \bibinfo{author}{\bibfnamefont{B.}~\bibnamefont{Hemmerling}},
  \bibinfo{author}{\bibfnamefont{E.}~\bibnamefont{Chae}},
  \bibinfo{author}{\bibfnamefont{J.~M.} \bibnamefont{Doyle}}, \bibnamefont{and}
  \bibinfo{author}{\bibfnamefont{J.}~\bibnamefont{Ye}}, \bibinfo{journal}{Phys.
  Rev. Lett.} \textbf{\bibinfo{volume}{114}}, \bibinfo{pages}{223003}
  (\bibinfo{year}{2015}).

\bibitem[{\citenamefont{Bernard et~al.}(1989)\citenamefont{Bernard, Effantin,
  d'Incan, Fabre, Stringat, and Barrow}}]{bernard}
\bibinfo{author}{\bibfnamefont{A.}~\bibnamefont{Bernard}},
  \bibinfo{author}{\bibfnamefont{C.}~\bibnamefont{Effantin}},
  \bibinfo{author}{\bibfnamefont{J.}~\bibnamefont{d'Incan}},
  \bibinfo{author}{\bibfnamefont{G.}~\bibnamefont{Fabre}},
  \bibinfo{author}{\bibfnamefont{R.}~\bibnamefont{Stringat}}, \bibnamefont{and}
  \bibinfo{author}{\bibfnamefont{R.}~\bibnamefont{Barrow}},
  \bibinfo{journal}{Mol. Phys.} \textbf{\bibinfo{volume}{67}},
  \bibinfo{pages}{1} (\bibinfo{year}{1989}).

\bibitem[{\citenamefont{Zhuang et~al.}(2011)\citenamefont{Zhuang, Le, Steimle,
  Bulleid, Smallman, Hendricks, Skoff, Hudson, Sauer, Hinds
  et~al.}}]{TarbuttZhuangPCCP11_FCFsYbF}
\bibinfo{author}{\bibfnamefont{X.}~\bibnamefont{Zhuang}},
  \bibinfo{author}{\bibfnamefont{A.}~\bibnamefont{Le}},
  \bibinfo{author}{\bibfnamefont{T.~C.} \bibnamefont{Steimle}},
  \bibinfo{author}{\bibfnamefont{N.~E.} \bibnamefont{Bulleid}},
  \bibinfo{author}{\bibfnamefont{I.~J.} \bibnamefont{Smallman}},
  \bibinfo{author}{\bibfnamefont{R.~J.} \bibnamefont{Hendricks}},
  \bibinfo{author}{\bibfnamefont{S.~M.} \bibnamefont{Skoff}},
  \bibinfo{author}{\bibfnamefont{J.~J.} \bibnamefont{Hudson}},
  \bibinfo{author}{\bibfnamefont{B.~E.} \bibnamefont{Sauer}},
  \bibinfo{author}{\bibfnamefont{E.~A.} \bibnamefont{Hinds}},
  \bibnamefont{et~al.}, \bibinfo{journal}{Phys. Chem. Chem. Phys.}
  \textbf{\bibinfo{volume}{13}}, \bibinfo{pages}{19013} (\bibinfo{year}{2011}).

\end{thebibliography}

\end{document}